\documentclass[aps,groupedaddress]{revtex4}
\usepackage{epsfig}
\usepackage{graphicx}
\usepackage[T1]{fontenc}
\usepackage{ae}
\usepackage{color}
\usepackage[latin1]{inputenc}
\usepackage{amssymb,amsbsy,amsmath}
\usepackage{bbm}

\newtheorem{prop}{Proposition}

\begin{document}



\title[CST]
      {Thermodynamics of the classical spin triangle}
\author{Heinz-J\"urgen Schmidt$^1$ and Christian Schr\"oder$^2$
}
\address{$^1$  Department of Physics,  Osnabr\"uck University,
 D - 49069 Osnabr\"uck, Germany\\
$^2$  Bielefeld Institute for Applied Materials Research, Bielefeld University of Applied Sciences, D - 33619 Bielefeld, Germany\\
and\\
Faculty of Physics, Bielefeld University, 33615 Bielefeld, Germany
}

\begin{abstract}
The classical spin system consisting of three spins with Heisenberg interaction is an example
of a completely integrable mechanical system. In this paper we explicitly calculate thermodynamic
quantities as density of states, specific heat, susceptibility and spin autocorrelation functions.
These calculations are performed (semi-)analytically and shown to agree with corresponding Monte Carlo simulations.
For the long-time autocorrelation function, we find, for certain values of the coupling constants,
a decay to constant values in the form of an $1/t$ damped harmonic oscillation and propose a theoretical explanation.
\end{abstract}

\maketitle

\section{Introduction}\label{sec:Intro}

Seemingly paradoxically, electronic quantum spin produces macroscopic magnetic effects described in classical terms.
An explanation of this is provided by the classical limit of quantum spin systems, see \cite{L73,FKL07}.
Many single spins with spin quantum number $s=1/2$ are combined into systems with larger $s$, integer or half-integer.
In the limit $s\to\infty$, the spin vector operator can be replaced by a classical spin vector of unit length after rescaling.
This limit can also be extended to systems with $N$ classical spins including their interaction.
The advantages of considering the classical limit are as follows:

\begin{itemize}
  \item The theory of classical spin systems, which is simpler than its quantum analogue, can be used to approximate the behavior of real systems
  with localized relatively large spins, e.g., of gadolinium atoms \cite{P15} with $s=7/2$ embedded in a magnetic molecule \cite{GOB14,QZYNSWZ17, QZYNSZ21}.
  \item The classical limit $s\to\infty$ can serve as a test for theoretical calculations performed for arbitrary $s$. Example,
  the high temperature expansion of specific heat or susceptibility leads to certain polynomials in $s$ whose leading coefficient can be calculated using classical theory, see e.~g.~\cite{SLR11}.
  \item A classical spin system can be defined as a system with $2N$-dimensional phase space
  ${\mathcal S}^2 \times \ldots \times {\mathcal S}^2$ and analyzed by the methods of classical mechanics. This procedure can be also applied to  the classical-statistical calculation of thermodynamic quantities.
\end{itemize}

The focus of the present work lies on the last item. We consider a classical ``spin triangle", i.~e., a spin system of $N=3$ spins with Heisenberg interaction
given by three generally different coupling constants $J_1, J_2, J_3$.
This is a phenomenological ansatz to describe the exchange interaction of spins by an isotropic Hamiltonian which is bi-linear
in the spin observables and has a straightforward classical analogue, see, e.~g., \cite[Eq.~(1.40)]{W15}. 
Real systems often show additional interactions, e.~g., of Dzyaloshinskii-Moriya type \cite{BRT18}, that complicate the theoretical treatment.
Every quantum spin system 
with Heisenberg Hamiltonian has three commuting observables that are constants of motion: The Hamiltonian itself, the square of the total
spin and its $3$-component. Since the corresponding classical functions on the six-dimensional phase space Poisson-commute we obtain a
\textit{completely integrable} classical system in the sense of the Arnol'd-Liouville theorem, see \cite{A78}.
Hence the time evolution can be explicitly calculated up to integrations, see \cite{S21}, in contrast to the situation for the quantum spin triangle \cite{S13}.
This is recapitulated in section \ref{sec:GD} with further details moved to Appendix \ref{sec:V}.

However, it is not self-evident that thermodynamic quantities such as density of states (dos), specific heat and susceptibility can also be calculated analytically for a completely integrable system. A positive example is given by the classical dipole pair \cite{SSHL15}. In the present case
we could not provide an expression for the dos that is valid for all choices of $J_1, J_2, J_3$ but suggest a procedure that has to be adapted
to every concrete case and give the details in Section \ref{sec:DS} for a standard example of $J_1, J_2, J_3$ used throughout the paper.
Calculating the specific heat in section \ref{sec:SH} using computer algebraic tools leads to expressions too complex to be presented in detail,
but which can be plotted and compared with Monte Carlo simulations.
Similarly, the zero field magnetic susceptibility can be calculated semi-analytically or numerically, see Section \ref{sec:SU}, with consistent results.

Another focus of the present work is the computation of the autocorrelation function (acf) for the general spin triangle, see Section \ref{sec:SA}.
This function has been frequently studied in the literature on classical spin systems
\cite{LLB98,CLAL99,MSL99,MSSL00,C00,AK02,C07,C22},
partly because it is of importance for proton spin-lattice relaxation
measurements  \cite{BM74,TM98,Letal16} and experiments with neutron scattering \cite{Betal12}.
The short time acf for low temperatures is characterized by peaks at the frequencies of spin waves that occur for energies
slightly above the ground state energy, see Section \ref{sec:STA}.

The spin triangle is treated in the literature mainly for the special cases of uniform coupling (equilateral triangle) or only
two different coupling constants (isosceles triangle and $3$-chain).
From the perspective of the current study of the general spin triangle, these special cases may create a somewhat biased impression:
For a large region in $J_1, J_2, J_3$-space  we find a decay of the acf to a constant value
in the form of a harmonic oscillation damped by the factor $1/t$, see Section \ref{sec:LTA},
while in the isosceles case other negative powers ($2$ or $3$) can occur, see \cite{AK02}.
Our finding of the $1/t$ decay is theoretically supported by the argument that the high-temperature peak of the Fourier-transformed acf
is a logarithmic singularity that occurs if the spin configurations possesses a saddle point
of the mean angular velocity of rotation, see appendices \ref{sec:SP} and \ref{sec:ALS}.
We close with a Summary in Section \ref{sec:SUM}.

Throughout this paper we will denote the dependence of functions $f(\ldots)$ on the 
(dimensionless) temperature $T$ and the inverse temperature $\beta=1/T$ by using the same letter $f$ without danger of confusion.

\section{General definitions and results concerning time evolution}\label{sec:GD}

We consider a classical spin system described by three spin vectors ${\mathbf s}_\mu,\;\mu=1,2,3,$ of unit length.
The corresponding six-dimensional phase space is
\begin{equation}\label{defphasespace}
  {\mathcal P}={\mathcal S}^2 \times {\mathcal S}^2 \times {\mathcal S}^2
  \;.
\end{equation}
Let $s$ denote the $3\times 3$-matrix with columns ${\mathbf s}_\mu,\;\mu=1,2,3$.
The total spin vector will be written as
\begin{equation}\label{deftotal}
{\mathbf S}:= {\mathbf s}_1+ {\mathbf s}_2+ {\mathbf s}_3
\;,
\end{equation}
and its length as $S$.
The general Heisenberg Hamiltonian $H: {\mathcal P}\rightarrow{\mathbbm R}$ will be written as
\begin{equation}\label{defHam}
 H(s)=J_1\, {\mathbf s}_2 \cdot  {\mathbf s}_3 + J_2\, {\mathbf s}_3 \cdot  {\mathbf s}_1 + J_3\, {\mathbf s}_1 \cdot  {\mathbf s}_2
 \;,
\end{equation}
with three real coupling coefficients $J_1, J_2, J_3$
and yields the corresponding Hamiltonian equations of motion, see \cite{S21},
\begin{eqnarray}\label{eom1}
  \dot{\mathbf s}_1&=& \left( J_2 \,{\mathbf s}_3+ J_3 \,{\mathbf s}_2\right)\times {\mathbf s}_1,\\
  \label{eom2}
  \dot{\mathbf s}_2&=& \left( J_3 \,{\mathbf s}_1+ J_1 \,{\mathbf s}_3\right)\times {\mathbf s}_2,\\
  \label{eom3}
  \dot{\mathbf s}_3&=& \left( J_1 \,{\mathbf s}_2+ J_2 \,{\mathbf s}_1\right)\times {\mathbf s}_3
  \;.
\end{eqnarray}
The equations of motion
(\ref{eom1} - \ref{eom3}) will be written in the compact form
\begin{equation}\label{eom4}
   \dot{s} = {\mathcal J}(s)
   \;,
\end{equation}
using the bilinear matrix-valued function ${\mathcal J}(s)$ with entries
\begin{equation}\label{defopJ}
 {\mathcal J}(s)_{i \mu}:=  \left( \sum_\kappa J_{\mu\kappa} {\mathbf s}_\kappa\times {\mathbf s}_\mu\right)^{(i)}
 \;,
\end{equation}
for $i,\mu=1,2,3$. Since the vector product transforms in a natural way under rotations $R\in SO(3)$ we have
\begin{equation}\label{RopJ}
  {\mathcal J}(R\,s)= R\, {\mathcal J}(s)
  \;,
\end{equation}
for all $R\in SO(3)$.

These equations of motion admit the conserved quantities $H(s), S^2(s), {\mathbf S}^{(3)}(s) $ assuming the values
\begin{eqnarray}
\label{conserved1}
  H(s) &=& \varepsilon\;, \\
  \label{conserved2}
  S^2(s) &=& 3+2\sigma \;,\\
  \label{conserved3}
 {\mathbf S}^{(3)}(s) &=& \sigma_3
 \;,
\end{eqnarray}
depending on the initial conditions, see, e.~g., \cite{S21}.

\begin{figure}[htp]
\centering
\includegraphics[width=0.7\linewidth]{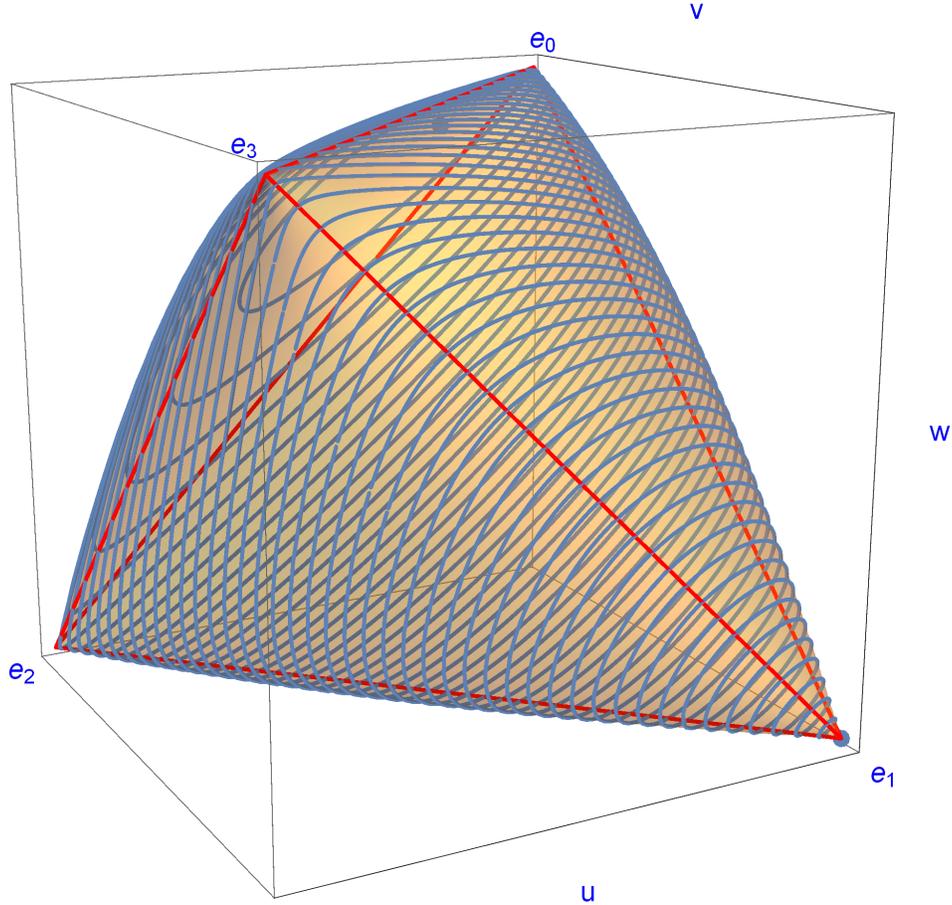}
\caption{Plot of the convex Gram set ${\mathcal G}$ defined in (\ref{defGram}).
${\mathcal G}$ has four singular extremal points denoted by ${\mathbf e}_i,\;i=0,1,2,3$.
The intersections of the planes $\widetilde{P}_\varepsilon$ with constant total energy $\varepsilon$ and the boundary $\partial{\mathcal G}$
are shown (blue curves) for $50$ different values of $\varepsilon$ and the choice
(\ref{choice}).
The minimal energy $E_{\scriptsize min}=-1-\sqrt{2}$ is assumed at ${\mathbf e}_1$ and the maximal
energy $E_{\scriptsize max}=\frac{1}{8}\left(10+\sqrt{2}\right)$ at the (blue) boundary point with coordinates (\ref{antiground}).
}
\label{FIGGRAM1}
\end{figure}

For the calculations concerning the thermodynamics of the system it is advisable to split the six degrees of freedom
into three internal variables and three external ones. As the internal variables we choose the three scalar products
\begin{equation}\label{defuvw}
  u:=  {\mathbf s}_2 \cdot  {\mathbf s}_3 ,\;
  v:= {\mathbf s}_3 \cdot  {\mathbf s}_1,\;
  w:= {\mathbf s}_1 \cdot  {\mathbf s}_2
  \;,
\end{equation}
together with the  scalar triple product
\begin{equation}\label{defdelta}
 \delta:= {\mathbf s}_1 \cdot\left(  {\mathbf s}_2 \times  {\mathbf s}_3  \right)=\pm \sqrt{1-u^2-v^2-w^2+2\,u\,v\,w}
 \;.
\end{equation}
The set of  vectors $(u,v,w)^\top\in {\mathbbm R}^3$ corresponding to spin configurations $s\in {\mathcal P}$ can be shown
\cite{SL03,S17a,S17b,S21}  to
form a convex set ${\mathcal G}$, called the "Gram set", defined by
\begin{equation}\label{defGram}
 {\mathcal G}:= \left\{ (u,v,w)^\top\in {\mathbbm R}^3 \left|\right. -1\le u,v,w\le 1\mbox{ and } 1-u^2-v^2-w^2+2\,u\,v\,w\ge 0\right\}
 \;,
\end{equation}
see Figure \ref{FIGGRAM1}.
The interior of ${\mathcal G}$, characterized by $ 1-u^2-v^2-w^2+2\,u\,v\,w > 0$, corresponds to three-dimensional spin configurations,
the boundary $\partial{\mathcal G}$, characterized by $ 1-u^2-v^2-w^2+2\,u\,v\,w = 0$, either corresponds to coplanar spin configurations
or to collinear ones, the latter generating the  four singular extremal points ${\mathbf e}_i,\;i=0,1,2,3,$ of ${\mathcal G}$,
corresponding to the spin configurations
\begin{equation}\label{singular}
 {\mathbf e}_0\widetilde{=}\uparrow\uparrow\uparrow,\; {\mathbf e}_1\widetilde{=}\downarrow\uparrow\uparrow,\;
{\mathbf e}_2\widetilde{=}\uparrow\downarrow\uparrow,\;{\mathbf e}_3\widetilde{=}\uparrow\uparrow\downarrow
\;,
\end{equation}
see Figure \ref{FIGGRAM1}.

The conserved quantities $H(s)=J_1 u +J_2 v+J_3 w=:E(u,v,w)$ and $S^2(s)=3+2\sigma=3+2(u+v+w)$ are linear functions of $u,v,w$.
Hence the equation  $H(s)=E(u,v,w)=\varepsilon$ defines a plane $\widetilde{P}_\varepsilon$ that intersects the Gram set
in a two-dimensional convex set $\widetilde{P}_\varepsilon\cap {\mathcal G}$, see Figure \ref{FIGGRAM1}. Exceptions are the extremal
values $\varepsilon=E_{\scriptsize min}$ and $\varepsilon=E_{\scriptsize max}$ where the intersection degenerates to a point.
Throughout this paper we will use the special choice
\begin{equation}\label{choice}
  J_1=-\frac{1}{2},\,J_2=\frac{1}{2}+\frac{\sqrt{2}}{2},\,J_3=\frac{\sqrt{2}}{2}
  \;,
\end{equation}
as a standard example. The same choice was made in \cite{S21} in order to simplify certain calculations in connection with the time evolution.
For this choice the minimal energy has the value $E_{\scriptsize min}=-1-\sqrt{2}$
which is assumed at ${\mathbf e}_1$ corresponding to a collinear ground state of the form $\downarrow\uparrow\uparrow$.
For this ground state each term of the Hamiltonian (\ref{defHam}) assumes its absolute minimum and hence the spin system is called ``unfrustrated".
In contrast, the anti-ground state where the energy $E_{\scriptsize max}=\frac{1}{8}\left(10+\sqrt{2}\right)$ is assumed
will be a coplanar state corresponding to a point ${\mathbf e}_4$ of the Gram set with coordinates
\begin{equation}\label{antiground}
 u_4=\frac{1}{4} \left(3 \sqrt{2}-2\right),\; v_4=2-\frac{3}{2 \sqrt{2}},\;w_4=\frac{1}{4} \left(3 \sqrt{2}-1\right)
 \;,
\end{equation}
see \cite{S21} for a proof. This coplanar state would be the ground state of a system with sign-inverted coupling constants (\ref{choice}).
For the latter system it is not possible to realize the absolute minimum of each term of the Hamiltonian (\ref{defHam})
and hence the system would be called ``frustrated".

Analogous considerations apply to the conservation law $S^2(s)=3+2(u+v+w)$ or, equivalently, $u+v+w=\sigma$.
Here the planes $P_\sigma$ are independent of the coupling constants and
perpendicular to the vector ${\mathbf e}_0=(1,1,1)^\top$. Again, ${P}_\sigma\cap {\mathcal G}$
will be a two-dimensional convex set, except for the extremal values $\sigma=-\frac{3}{2}$ and $\sigma=3$,
corresponding to $S^2=0$ and $S^2=9$, resp.~.

The intersection of the two planes $P_\sigma$ and $\widetilde{P}_\varepsilon$ forms a line $L(\sigma,\varepsilon)$.
On this line $v$ and $w$ can be expressed through $u$ via
\begin{equation}\label{u2v}
  v=\frac{J_3 (u-\sigma )-J_1 u+\varepsilon }{J_2-J_3}
  \;,
\end{equation}
and
\begin{equation}\label{u2w}
  w=\frac{-J_2 (u-\sigma )+J_1 u-\varepsilon }{J_2-J_3}
  \;.
\end{equation}

The time evolution of the internal variables turns out to be an oscillation along the bounded interval
$L(\varepsilon,\sigma)\cap {\mathcal G}$ such that each point is traversed in two directions. To make the
time evolution unique, we will replace ${\mathcal G}$ by its ``double"
\begin{equation}\label{defGprime}
 {\mathcal G}':=\{(u,v,w,\delta)\in{\mathbbm R}^4\left| \right. \delta^2=1-(u^2+v^2+w^2)+2\,u\,v\,w\;\text{ and } -1\le u,v,w\le 1,
 \mbox{ but not }\left| u\right| =\left| v\right| =\left| w\right| =1\}
 \;,
\end{equation}
where we have further excluded the singular extremal points.
${\mathcal G}'$ essentially consists of two copies of ${\mathcal G}$ glued together at their boundaries
and will be a three-dimensional manifold, see \cite{S21}.
The equations of motion (\ref{eom1} - \ref{eom3}) give rise to the following autonomous system of differential equations
for the internal variables:
\begin{eqnarray}
\label{udot}
  \dot{u} &=& \left( J_3-J_2\right)\,\delta, \\
  \label{vdot}
   \dot{v} &=& \left( J_1-J_3\right)\,\delta, \\
  \label{wdot}
   \dot{w} &=& \left( J_2-J_1\right)\,\delta,\\
  \label{deltadot}
\dot{\delta}&=& J_1\,(u+1)(w-v)+J_2\,(v+1)(u-w)+J_3\,(w+1)(v-u)
\;.
\end{eqnarray}
According to \cite{S21} the solution of (\ref{udot} - \ref{wdot}) can be written as
\begin{eqnarray}
\label{uvwt1}
  u(t) &=& \frac{1}{g}\left({\wp \left(t+t_0;g_2,g_3\right)-x_0} \right),\\
  \label{uvwt2}
  v(t) &=& \frac{1}{g \left(J_2-J_3\right)}
 \left( \left(J_3-J_1\right) \wp \left(t+t_0;g_2,g_3\right)-J_3 \left(g \sigma +x_0\right)+g \varepsilon +J_1   x_0
 \right),\\
 \label{uvwt3}
 w(t)&=&\frac{1}{g \left(J_2-J_3\right)}
 \left(\left(J_1-J_2\right) \wp \left(t+t_0;g_2,g_3\right)+J_2\left(g \sigma + x_0\right)-g \varepsilon -J_1 x_0
 \right)
 \;,
\end{eqnarray}
where $\wp \left(z;g_2,g_3\right)$ denotes the \textit{Weierstrass elliptic function}, see, e.~g.,  \cite[Ch.23]{NIST21},
and the parameters $g,x_0, g_2,g_3, t_0$ are given in Appendix \ref{sec:V} as functions of $\varepsilon, \sigma, J_1,J_2,J_3$.
The solution (\ref{uvwt1} - \ref{uvwt3}) is ${\sf T}$-periodic with ${\sf T}={\sf T}(\sigma,\varepsilon)$ given in Appendix \ref{sec:V}.
By inserting (\ref{uvwt1} - \ref{uvwt3}) into $\delta(t)=\pm\sqrt{1-u^2(t)-v^2(t)-w^2(t)+2\,u(t)\,v(t)\,w(t)}$
and choosing the sign $\pm$ such that $t\mapsto \delta(t)$ will be a smooth function
we can also obtain a solution of (\ref{deltadot}) for all $t\in {\mathbbm R}$.

For each point $(u,v,w,\delta)\in{\mathcal G}'$ (except $u=v=w=-1/2$) we define the ``standard  spin configuration"
$r(u,v,w,\delta)=\left({\mathbf r}_1,{\mathbf r}_2,{\mathbf r}_3\right)\in{\mathcal P}$ that realizes the internal variables
$(u,v,w,\delta)$ and leads to a total spin vector ${\mathbf R }={\mathbf r}_1+{\mathbf r}_2+{\mathbf r}_3=(0,0,\sqrt{3+2\sigma})^\top$, see \cite{S21}:
\begin{equation}\label{defr1}
 {\mathbf r}_1=\left(
\begin{array}{c}
 \sqrt{\frac{2 (u+1)-(v+w)^2}{3+2( u+ v+ w)}} \\
 0 \\
 \frac{v+w+1}{\sqrt{3+2( u+ v+ w)}} \\
\end{array}
\right)\;,
\end{equation}

\begin{equation}\label{defr2}
 {\mathbf r}_2=\left(
\begin{array}{c}
 \frac{w (u+v+1)-(u+1) (v+1)+w^2}{ \sqrt{(2 (u+1)-(v+w)^2)(3+2( u+ v+ w)})} \\
 \frac{\delta}{\sqrt{2(u+1)-(v+w)^2}} \\
 \frac{u+w+1}{\sqrt{3+2 (u+v+w)}} \\
\end{array}
\right)\;,
\end{equation}

\begin{equation}\label{defr3}
 {\mathbf r}_3=\left(
\begin{array}{c}
 \frac{v (w+u+1)-(w+1) (u+1)+v^2}{ \sqrt{(2 (u+1)-(v+w)^2)(3+2( u+ v+ w)})} \\
 - \frac{\delta}{\sqrt{2(u+1)-(v+w)^2}} \\
 \frac{u+v+1}{\sqrt{3+2 (u+v+w)}} \\
\end{array}
\right)\;.
\end{equation}

An arbitrary spin configuration $s$  realizing the internal variables $(u,v,w,\delta)\in{\mathcal G}'$ can be written as
\begin{equation}\label{r2s}
  s= R\,r(u,v,w,\delta)
  \;,
\end{equation}
where $R$ is a unique proper rotation $R\in SO(3)$. It is hence sensible to define the three parameters determining $R$,
for example, the three Euler angles $\alpha,\beta,\gamma$ parametrizing $R$, as the {\it external} variables of the spin configuration.

By inserting the solutions (\ref{uvwt1} - \ref{uvwt3}) into $r(u,v,w,\delta)$ we obtain a time-dependent configuration
$r(t)=r(u(t),v(t),w(t),\delta(t)$. Although $r(t)$ does, in general, not satisfy the equations of motion (\ref{eom1} - \ref{eom3})
it can be shown that the corresponding solution $s(t)$ with initial condition $s(0)=r(0)$ is of the form
\begin{equation}\label{solst}
 s(t) =Z(t)\,r(t)
 \;,
\end{equation}
where $Z(t)\in SO(3)$ and $Z(t)\,{\mathbf R }={\mathbf R }$. Hence
\begin{equation}\label{formZ}
 Z(t)= \left(
\begin{array}{ccc}
 \cos \alpha(t)&-\sin \alpha(t) & 0 \\
 \sin \alpha(t) & \cos \alpha(t) & 0 \\
 0& 0 & 1 \\
\end{array}
\right)
\;.
 \end{equation}
  (\ref{formZ}) implies
 \begin{equation}\label{Zt}
   \frac{d}{dt}Z(t)=\Omega(t)\, Z(t)= Z(t)\,  \Omega(t)
   \;,
 \end{equation}
 where $\Omega(t)$ is the anti-symmetric ``angular velocity  matrix"
 \begin{equation}\label{defOmega}
  \Omega(t)= \dot{\alpha}(t)\,\left(
\begin{array}{ccc}
0&-1 & 0 \\
1 & 0 & 0 \\
 0& 0 &0 \\
\end{array}
\right)
\;.
 \end{equation}
 Hence
 \begin{equation}\label{sdot1}
   \dot{s}\stackrel{(\ref{solst})}{=}\dot{Z}r+Z\dot{r}\stackrel{(\ref{Zt})}{=}Z\left( \Omega\,r+\dot{r}\right)
   \;,
 \end{equation}
and further
\begin{equation}\label{sdotminusJ}
   \dot{s}-{\mathcal J}(s)\stackrel{(\ref{sdot1},\ref{solst})}{=}Z\left( \Omega\,r+\dot{r}\right)-{\mathcal J}(Z r)
   \stackrel{(\ref{RopJ})}{=}Z\left( \Omega\,r+\dot{r}-{\mathcal J}(r)\right)
   \;.
\end{equation}
Thus the equation of motion $0= \dot{s}-{\mathcal J}(s)$ is equivalent to
\begin{equation}\label{eomext}
 0= \Omega\,r +\dot{r}-{\mathcal J}(r)
 \;.
\end{equation}
In the case of an invertible $r$ the solution of (\ref{eomext}) is given by
\begin{equation}\label{soleomext}
 \Omega(t)= \left( {\mathcal J}(r(t))-\dot{r}(t)\right) r^{-1}(t)
 \;,
\end{equation}
and can be extended to $t$ being an integer multiple of $\frac{\sf T}{2}$,
where $r(t)$ is coplanar and hence not invertible, by means of continuity,
see \cite{S21} for the details.
The rotation matrix $Z(t)$ is then given by an integral over $t$:
Taking into account the form of $\Omega(t)$ according to (\ref{defOmega}) we obtain
\begin{equation}\label{intalpha}
 \alpha(t)=\int_{0}^{t}\dot{\alpha}(t')\,dt'=\int_{0}^{t}\left(  \left( {\mathcal J}(r(t'))  -\dot{r}(t')\right) \, r(t')^{-1}\right)_{2,1}\,dt'
 \;,
\end{equation}

\section{Density of states}\label{sec:DS}

\begin{figure}[htp]
\centering
\includegraphics[width=0.5\linewidth]{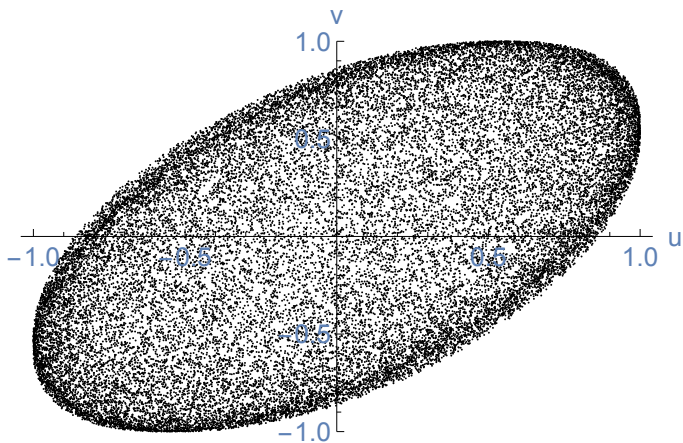}
\caption{Plot of the internal variables $(u,v)$ for approximately $30.000$  randomly chosen spin configurations selected
according to $0.5<w<0.6$.
The random choice of the spin configuration is performed according to a uniform probability distribution w.~r.~t.~the
canonical coordinates (\ref{defcan}).
Hence this plot can be viewed as a cut of the density of states function in the Gram set.
Obviously, the density concentrates at the boundary of the Gam set according to the factor $1/\delta$ in (\ref{newdV}).
}
\label{FIGGRA}
\end{figure}

In the first step we want to calculate the density of states in the Gram set ${\mathcal G}$.
The volume element $dV$ of ${\mathcal P}$ can be written as
\begin{equation}\label{dV}
 dV= d\phi_1\, dz_1\, d\phi_2\, dz_2\, d\phi_3\, dz_3\;,
\end{equation}
using the ``canonical coordinates" $\left(\phi_\mu,\,z_\mu\right)_{\mu=1,2,3}$ defined
via the representation
\begin{equation}\label{defcan}
  {\mathbf s}_\mu =
  \left(
\begin{array}{c}
 \sqrt{1-z_{\mu }^2} \cos\,\phi _{\mu } \\
 \sqrt{1-z_{\mu }^2} \sin \,\phi _{\mu } \\
 z_{\mu } \\
\end{array}
\right)
\;.
\end{equation}
We will pass from canonical coordinates $\left(\phi_\mu,\,z_\mu\right)_{\mu=1,2,3}$ to the
coordinates $(\alpha,\beta,\gamma, u,v,w)$ defined by
\begin{equation}\label{defnewcoor}
 s=R(\alpha,\beta,\gamma)\,r(u,v,w,\delta)
 \;,
\end{equation}
where $R(\alpha,\beta,\gamma)$ denotes the parametrization of a rotation by Euler angles.
After some calculations the volume element in terms of the new coordinates is obtained as
\begin{equation}\label{newdV}
 dV = \frac{\partial(\phi_1, z_1,\phi_2,z_2,\phi_3,z_3)}{\partial(\alpha,\beta,\gamma,u\,v\,w)} \,d\alpha \,d\beta\,d\gamma\,du\,dv\,dw
 = \frac{\sin \beta}{\delta}\,d\alpha \,d\beta\,d\gamma\,du\,dv\,dw
 \;.
\end{equation}
Integration of (\ref{newdV}) over $d\alpha \,d\beta\,d\gamma$ yields the volume of $SO(3)$. This gives a constant factor that can be neglected
since it finally cancels anyway when dividing by the partition function. The remaining volume element of ${\mathcal G}$ will thus be
\begin{equation}\label{dVp1}
  dV'= \frac{1}{\delta}\,du\,dv\,dw
  \;.
\end{equation}
The essential message is that the density of states in the Gram set
is {\em not} uniform but proportional to $1/\delta$ and thus diverging at the boundary of ${\mathcal G}$, see Figure \ref{FIGGRA}.\\

Next we will consider the density of states function $D(\sigma,\varepsilon)$ that is needed for
calculating the susceptibility. To this end we pass from the coordinates $(u,v,w)$ of the Gram set
to new coordinates $(\sigma,\varepsilon,t)$. The coordinate $t$ has to be understood as follows: For fixed $\sigma$ and $\varepsilon$
the system moves on the interval $L(\sigma,\varepsilon)\cap {\mathcal G}$ according to (\ref{uvwt1}-\ref{uvwt3}). For this motion
and $0< t <{\sf T}/2$ the internal variable $u(t)$ will be a smooth $1:1$ function of $t$, if $t=0$ is chosen as the point in time where
the system passes the boundary of ${\mathcal G}$. Hence $t$ can be used as a coordinate
parametrizing $L(\sigma,\varepsilon)\cap {\mathcal G}$. After some calculations using (\ref{udot}) we obtain the Jacobian
\begin{equation}\label{jacuvw}
  \frac{\partial(u,v,w)}{\partial(\sigma, \varepsilon,t)} =\delta
  \;,
\end{equation}
and hence
\begin{equation}\label{dVp}
 dV'\stackrel{(\ref{dVp1})}{=} \frac{1}{\delta}\,du\,dv\,dw
 = \frac{1}{\delta}   \frac{\partial(u,v,w)}{\partial(\sigma, \varepsilon,t)}\,d\sigma\,d\varepsilon\,dt
 \stackrel{(\ref{jacuvw})}{=} d\sigma\, d\varepsilon\,dt
 \;.
\end{equation}
This result means that the density of states restricted to the interval $L(\sigma,\varepsilon)\cap {\mathcal G}$
can be obtained from the time average of occupying an interval $(u, u+du)$ and thus reminds us of the definition of ergodicity.
The slower the system moves, the higher is its density of states and consequently the latter diverges at the endpoints of the
interval $L(\sigma,\varepsilon)\cap {\mathcal G}$ where $\dot{u}=(J_3-J_2)\delta=0$.

Integrating over $t$ yields
\begin{equation}\label{Dse}
    dV'':=2\int_{0}^{{\sf T}/2}dV'\stackrel{(\ref{dVp})}{=}2\int_{0}^{{\sf T}/2}  d\sigma\, d\varepsilon\,dt =
    {\sf T}(\sigma,\varepsilon)\,d\sigma\, d\varepsilon
     =:D(\sigma,\varepsilon)\,d\sigma\,d\varepsilon
\;,
\end{equation}
where we have inserted the irrelevant factor $2$ for the sake of simplicity.
The density of states function $D(\sigma,\varepsilon)$ can thus be identified with the known function ${\sf T}(\sigma,\varepsilon)$
but its explicit dependence on $(\sigma,\varepsilon)$ would be rather complicated.
The domain of definition of  $D(\sigma,\varepsilon)$ will be the set
\begin{equation}\label{defSigma}
 \Sigma =\left\{(\sigma,\varepsilon)\left| E_{\scriptsize min}\le \varepsilon\le E_{\scriptsize max}
 \mbox{ and } \sigma_{\scriptsize min}(\varepsilon)\le \sigma \le \sigma_{\scriptsize max}(\varepsilon) \right.\right\}
 \;,
\end{equation}
where the bounds $\sigma_{\scriptsize min}(\varepsilon)$ and $\sigma_{\scriptsize max}(\varepsilon)$ cannot be expressed
explicitly but as the roots of a polynomial of order six.

For the specific heat we would need the pure energy-depending marginal density of states function $D(\varepsilon)$, again denoted by the same
letter without danger of confusion. Direct integration of $D(\sigma,\varepsilon)$  over $\sigma$ seems to be very difficult due to the complicated
dependence on $\sigma$. Hence we will pursue another way. Recall that the intersection $\widetilde{P}_\varepsilon\cap {\mathcal G}$
of the constant energy plane $\widetilde{P}_\varepsilon$ with the Gram set will be a  two-dimensional convex set that can be parametrized
by $(u,v)$, see Figure \ref{FIGGRAM1}.
We will determine $D(\varepsilon)$ by integrating $dV'$ given by (\ref{dVp1}) over $\widetilde{P}_\varepsilon\cap {\mathcal G}$.
The results of these integrations depend on a case distinction related to four energy intervals in the following way:
For every spin triangle there exist five critical energies, namely the four energies $E({\mathbf e}_i),\,i=0,\ldots,3$
at the singular extremal points of ${\mathcal G}$, together with  $E({\mathbf e}_4)$, where ${\mathbf e}_4=(u_4,v_4,w_4)$
is given by
\begin{eqnarray}
\label{e4u}
 u_4 &=&  \frac{1}{2}\left( \frac{J_2 J_3}{J_1^2}-\frac{J_2}{J_3}-\frac{J_3}{J_2}\right), \\
 \label{e4v}
 v_4 &=& \frac{1}{2}\left( \frac{J_1 J_3}{J_2^2}-\frac{J_1}{J_3}-\frac{J_3}{J_1}\right),\\
 \label{e4w}
 w_4 &=& \frac{1}{2}\left( \frac{J_1 J_2}{J_3^2}-\frac{J_1}{J_2}-\frac{J_2}{J_1}\right)
 \;,
\end{eqnarray}
and hence
\begin{equation}\label{E4}
  E({\mathbf e}_4)=J_1\,u_4+J_2\,v_4+J_3\,w_4= J_1+J_2+J_3 -\frac{\left(J_1 J_2+J_3 J_2+J_1 J_3\right){}^2}{2 J_1 J_2 J_3}
  \;,
\end{equation}
see \cite{S21}.
$ E({\mathbf e}_4)$ will always be either the ground state energy $E_{\scriptsize min}$ or the anti-ground state energy $E_{\scriptsize max}$.
If the five energies $E({\mathbf e}_i),\,i=0,\ldots,4$ are linearly ordered we obtain
four energy intervals with endpoints $E({\mathbf e}_i),\,i=0,\ldots,4$. This order depends on the $J_1,J_2,J_3$ and hence
influences the results of the above-mentioned integrations.\\

\begin{figure}[htp]
\centering
\includegraphics[width=0.5\linewidth]{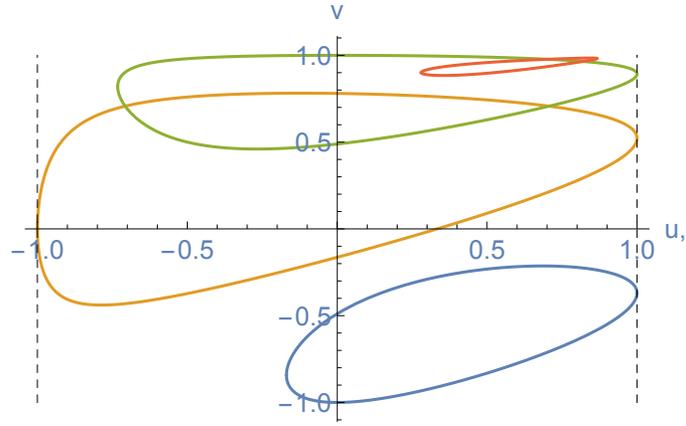}
\caption{Plot of the intersection $\widetilde{P}_\varepsilon \cap \partial{\mathcal G}$ for the standard choice of coupling constants (\ref{choice})
and the four values
$\varepsilon=\frac{1}{2} \left(-1-\sqrt{2}\right),\frac{1}{2},\frac{1}{2} \left(1+\sqrt{2}\right),\frac{1}{2}
   \left(\sqrt{2}+\frac{1}{8} \left(10+\sqrt{2}\right)\right)$. These values are examples for the four cases (\ref{ulim1} - \ref{ulim4}) and
   correspond to the four (blue, yellow, green red) closed curves, resp.~.
   }
\label{FIGPP}
\end{figure}

\begin{figure}[htp]
\centering
\includegraphics[width=0.95\linewidth]{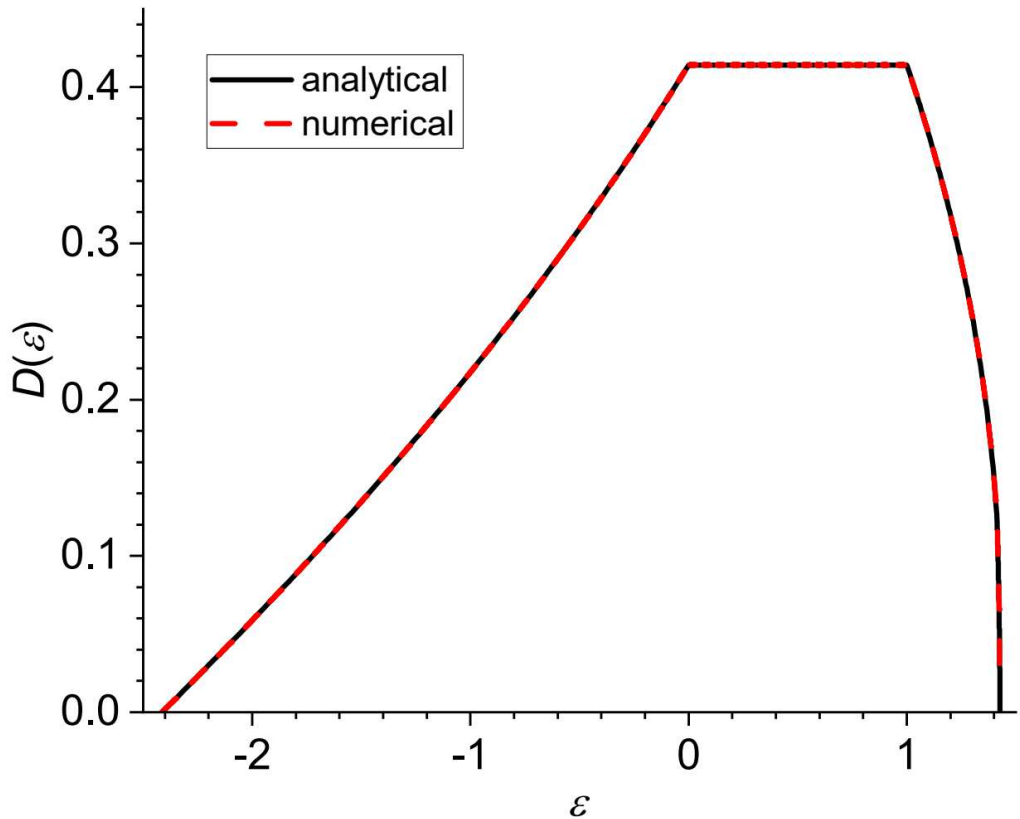}
\caption{Comparison of the analytical density of states function
$D(\varepsilon)$ according to (\ref{Depsint}) with the numerical result obtained by Wang-Landau sampling \cite{WL01}.
   }
\label{FIGDOS1}
\end{figure}

According to the previous remarks it will be advisable to fix a choice of $J_1,J_2,J_3$ in order to explain the steps
towards calculating $D(\varepsilon)$ having in mind that a different choice would lead to different but analogous equations.
We will choose the values (\ref{choice}) of the standard example used in this paper and obtain the critical energies
\begin{eqnarray}\label{Ecritex0}
E_0&=&E({\mathbf e}_1)=-1-\sqrt{2}=E_{\scriptsize min},\\
\label{Ecritex1}
E_1&=&E({\mathbf e}_3)=0,\\
\label{Ecritex2}
E_2&=&E({\mathbf e}_2)=1,\\
\label{Ecritex3}
E_3&=&E({\mathbf e}_0)=\sqrt{2},\\
\label{Ecritex4}
E_4&=&E({\mathbf e}_4)=\frac{1}{8} \left(10+\sqrt{2}\right)=E_{\scriptsize max}.
\end{eqnarray}
For each value of $\varepsilon$ the closed curve $\widetilde{P}_\varepsilon \cap \partial{\mathcal G}$
is given by the graph of the two functions
\begin{equation}\label{u2vpm}
  v_\pm(u)=\frac{\left(u \left(\sqrt{2} u+2 \sqrt{2} \varepsilon +\sqrt{2}+1\right)\pm \sqrt{2} \sqrt{\left(u^2-1\right)
   \left(u^2-2 u \left(-2 \varepsilon +\sqrt{2}+2\right)+4 \varepsilon ^2-2 \sqrt{2}-5\right)}\right)+2
   \left(1+\sqrt{2}\right) \varepsilon }{2 \left(2+\sqrt{2}\right) u+2 \sqrt{2}+5}
   \;,
\end{equation}
where $u$ varies between two extremal values
\begin{equation}\label{upm}
  u_\pm =-\left(2 \varepsilon \pm \sqrt{-4 \left(2+\sqrt{2}\right) \varepsilon +6 \sqrt{2}+11}\right)+\sqrt{2}+2
  \;,
\end{equation}
except for the cases where $\left| u_\pm \right|>1$. More precisely, the limits $u_1\le u \le u_2$ depend on the energy interval
in the following way:
\begin{eqnarray}
\label{ulim1}
 E_0\le \varepsilon \le E_1 &\Rightarrow& u_- \le u \le 1\;, \\
\label{ulim2}
 E_1\le \varepsilon \le E_2 &\Rightarrow& -1 \le u \le 1\;, \\
 \label{ulim3}
 E_2\le \varepsilon \le E_3 &\Rightarrow& u_- \le u \le 1\;, \\
 \label{ulim4}
 E_3\le \varepsilon \le E_4 &\Rightarrow& u_- \le u \le  u_+\;,
\end{eqnarray}
see Figure \ref{FIGPP}. Then it is a straightforward task to obtain $D(\varepsilon)$ by the integration
\begin{equation}\label{Deps}
 D(\varepsilon) \sim \int_{u_1}^{u_2}\int_{v_-(u)}^{v_+(u)}\frac{1}{\delta}\,dv\,du
 \;,
\end{equation}
where an overall factor has been left open. If this factor is chosen such that normalization
$\int_\varepsilon D(\varepsilon)\,d\varepsilon=1$ is achieved
the result reads as follows:
\begin{equation}\label{Depsint}
 D(\varepsilon)=
 \left\{
\begin{array}{l@{\;:\;}l}
 \frac{3}{2 \sqrt{2}}-\frac{\sqrt{\frac{1}{8} \left(10+\sqrt{2}\right)-\varepsilon
   }}{\sqrt{2+\sqrt{2}}}&E_0\le \varepsilon \le E_1,\\
\sqrt{2}-1 &E_1\le \varepsilon \le E_2,\\
\frac{\sqrt{\frac{1}{8} \left(10+\sqrt{2}\right)-\varepsilon }}{\sqrt{2+\sqrt{2}}}+\frac{3}{2
   \sqrt{2}}-1& E_2\le \varepsilon \le E_3 ,\\
\sqrt{4-2 \sqrt{2}} \sqrt{\frac{1}{8} \left(10+\sqrt{2}\right)-\varepsilon }
&E_3\le \varepsilon \le E_4
\;,
\end{array}
\right.
\end{equation}
see Figure \ref{FIGDOS1}. Remarkably, there exists a plateau of $D(\varepsilon)$ for $E_1\le \varepsilon\le E_2$,
similarly as in the case of a pair of dipoles \cite{SSHL15}.

\begin{figure}[htp]
\centering
\includegraphics[width=0.7\linewidth]{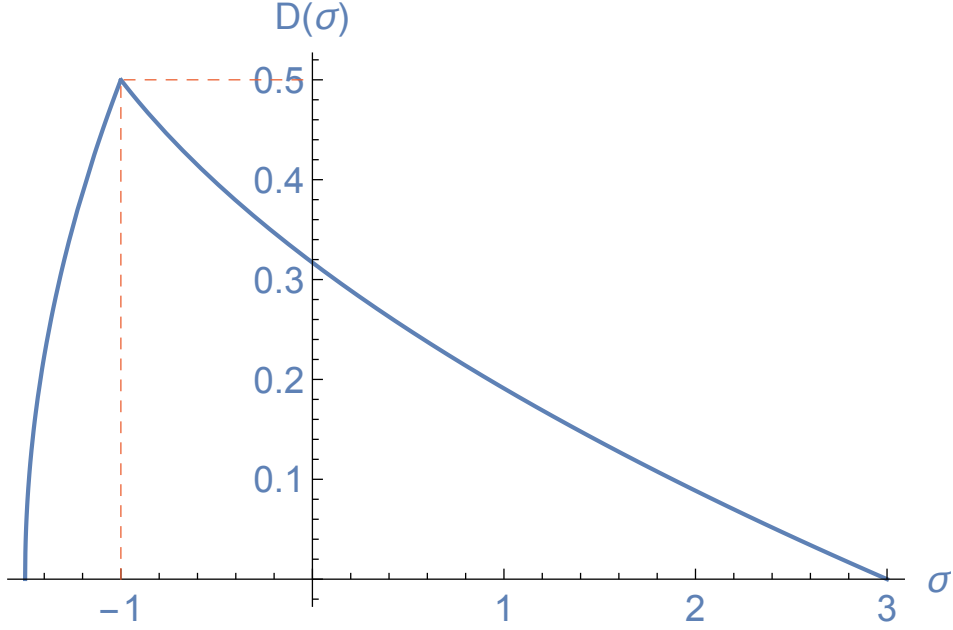}
\caption{Plot of the density of states function $D(\sigma)$ defined for $-3/2\le \sigma \le 3$, see (\ref{dsigma1}).
Its maximum of $1/2$ is attained at $\sigma=-1$.
   }
\label{FIGSIG}
\end{figure}
We will also consider the other marginal density function obtained by
\begin{equation}\label{defDsigma}
 D(\sigma)=\int_{\Sigma} D(\sigma,\varepsilon)\,d\varepsilon
 \;,
\end{equation}
again denoted by the same letter without danger of confusion. Again, direct integration of $D(\sigma,\varepsilon)$
would be too difficult but an analogous procedure as used for $D(\varepsilon)$ leads to the (normalized) result
\begin{equation}\label{dsigma1}
 D(\sigma)= \left\{
\begin{array}{l@{\;:\;}l}
\frac{1}{2 \sqrt{2}}\,\left(\sqrt{\sigma +\sqrt{2 \sigma +3}+2}-\sqrt{\sigma -\sqrt{2 \sigma +3}+2}\right)
&-3/2\le \sigma \le -1,\\
\frac{1}{4} \left(2-\sqrt{2} \sqrt{\sigma -\sqrt{2 \sigma +3}+2}\right)
&-1\le \sigma \le 3
\;,
\end{array}
\right.
\end{equation}
see Figure \ref{FIGSIG}.
From (\ref{dsigma1}) it follows by a short calculation that the mean value of $\sigma$ vanishes:
\begin{equation}\label{sigma0}
  \int_{-3/2}^{3}\sigma\,D(\sigma)\,d\sigma=0
  \;.
\end{equation}
Alternatively, (\ref{sigma0}) can be proven by viewing $\sigma$
as the energy of a spin triangle with $J_1=J_2=J_3=1$
and anticipating (\ref{Uvanish}).

\section{Specific heat}\label{sec:SH}

\begin{figure}[htp]
\centering
\includegraphics[width=1.0\linewidth]{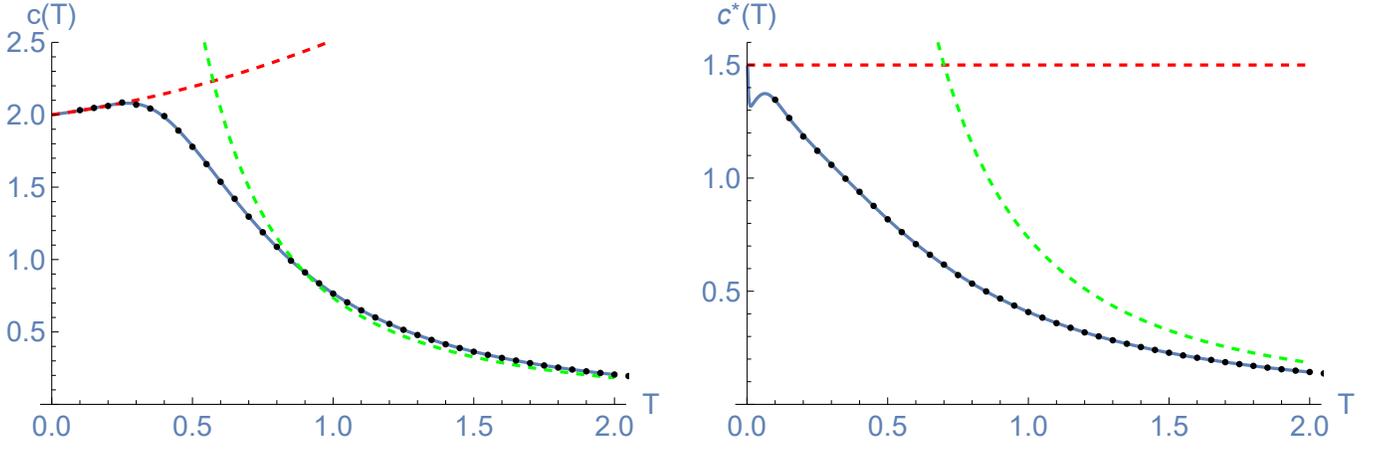}
\caption{The specific heat $c(T)$ for the spin triangle with coupling constants (\ref{choice}) (left panel) as well
as  $c^\ast(T)$ for the sign-inverted system (right panel). The analytical results are shown as blue curves; the numerical
ones as back points. We have also indicated the low-temperature limits
(\ref{lTlim1}) and (\ref{lTlim2}) (dashed, red curves) and the
high-temperature one (\ref{hTlim}) (dashed, green curves).
   }
\label{FIGC12}
\end{figure}

\begin{figure}[htp]
\centering
\includegraphics[width=1.0\linewidth]{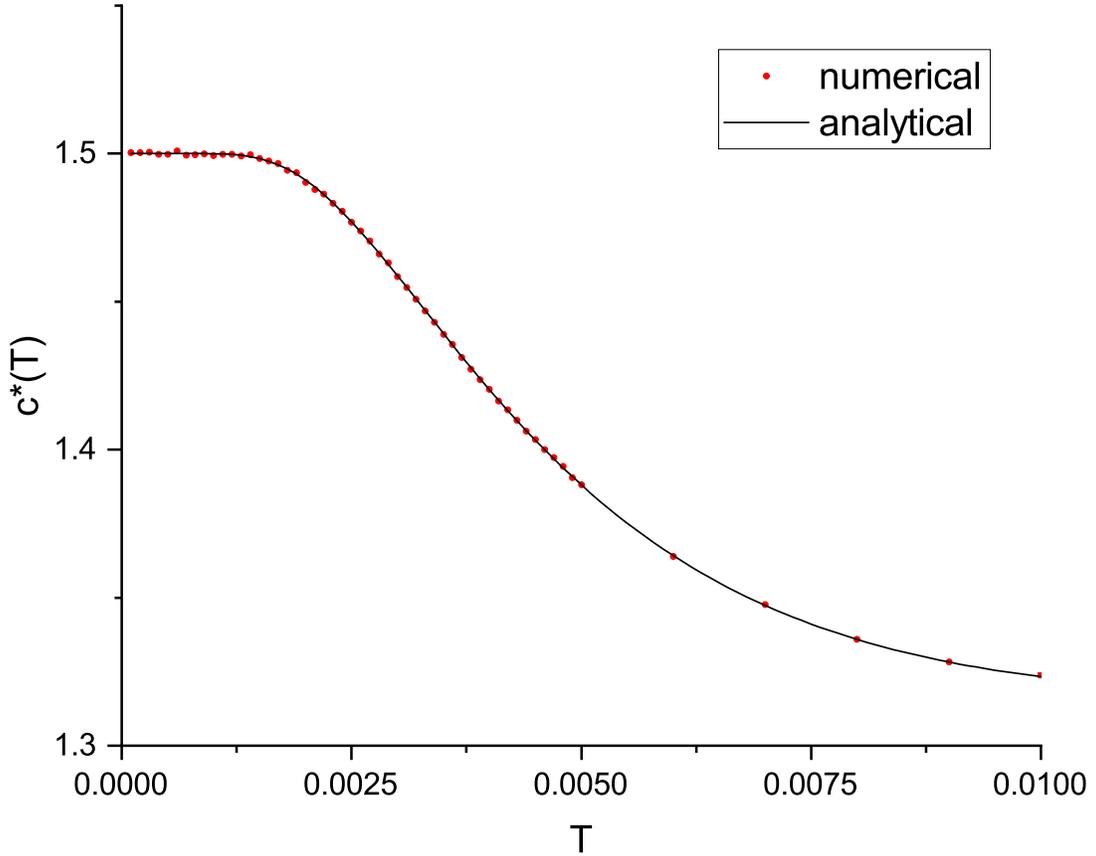}
\caption{The specific heat $c^\ast(T)$ for the spin triangle
with sign-inverted coupling constants (\ref{choice}) restricted to low temperatures
such that the almost constant value of $c^\ast(T)\approx 1.5$ for $0\le T\lesssim 0.001$ becomes visible.
The analytical result (black curve) and the numerical one (red dots) agree perfectly.
}
\label{FIGC1}
\end{figure}

For the calculation of the specific heat we do not have to resort to the phase space, but can
use the density-of-states function $D(\varepsilon)$ modulated by the Boltzmann factor $e^{-\beta\,\varepsilon}$,
where $\beta=1/T$ denotes the (dimensionless) inverse temperature, as usual.
Then the  partition function can be defined as
\begin{equation}\label{defZ}
 Z(\beta)= \int_{E_{\scriptsize min}}^{E_{\scriptsize max}} D(\varepsilon)\,e^{-\beta\,\varepsilon}\,d\varepsilon
 \;,
\end{equation}
and the inner energy as
\begin{equation}\label{defU1}
U_1(\beta)=\frac{1}{Z(\beta)}\,\int_{E_{\scriptsize min}}^{E_{\scriptsize max}}\varepsilon\, D(\varepsilon)\,e^{-\beta\,\varepsilon}\,d\varepsilon
\;.
\end{equation}
For later use we note that the inner energy vanishes in the high temperature limit:
\begin{equation}\label{Uvanish}
U_1(\beta=0)=\frac{1}{Z(\beta=0)}\,\int_{E_{\scriptsize min}}^{E_{\scriptsize max}}\varepsilon\, D(\varepsilon)\,d\varepsilon=0
\;.
\end{equation}
This holds generally for Heisenberg couplings since $\left\langle {\mathbf s}_\mu\right\rangle_\beta={\mathbf 0}$ and
$\left\langle {\mathbf s}_\mu\cdot{\mathbf s}_\nu\right\rangle_\beta=\left\langle {\mathbf s}_\mu\right\rangle_\beta
\cdot\left\langle{\mathbf s}_\nu\right\rangle_\beta=0$ for all $\mu,\nu$ and $\beta=0$.

Together with the second moment
\begin{equation}\label{defU2}
U_2(\beta)=\frac{1}{Z(\beta)}\,\int_{E_{\scriptsize min}}^{E_{\scriptsize max}}\varepsilon^2\, D(\varepsilon)\,e^{-\beta\,\varepsilon}\,d\varepsilon
\;,
\end{equation}
after a short calculation the specific heat $c(T):=\frac{\partial}{\partial T} U_1(1/T)$ can be obtained in the well-known form as
\begin{equation}\label{defc}
  c(\beta)=\beta^2 \left( U_2(\beta)-U_1^2(\beta)\right)
  \;.
\end{equation}

For the spin triangle the integrals involved in the calculation of $c(\beta)$ can be analytically calculated with the aid
of computer-algebraic software but the result is usually too complicated to be explicitly presented.
Nevertheless, we will plot the specific heat for the standard example (\ref{choice}) and compare it with numerical calculations.

To each spin triangle with coupling constants $J_1,J_2,J_3$ there belongs a sign-inverted system with coupling
constants $J^\ast_i=-J_i$ for $i=1,2,3$. Obviously, its density of state function is obtained as $D^\ast(\varepsilon)=D(-\varepsilon)$
and the specific heat $c^\ast(\beta) = c(-\beta)$ by extending the original function $c(\beta)$ to negative arguments.
Using this fact we will display the specific heat $c(T)$ for the (un-frustrated) example (\ref{choice}) as well as $c^\ast(T)$
for the (frustrated) sign-inverted system, see Figure \ref{FIGC12}.

The high-temperature limit of $c(T)$ is given by the variance of $\varepsilon$, calculated for $\beta=0$, divided by $T^2$.
It is always the same for the system and the sign-inverted one.
For our example we obtain
\begin{equation}\label{hTlim}
  c(T) = c^\ast(T)=\frac{3+\sqrt{2}}{6\, T^2}+O(T^{-3})
  ;,
\end{equation}
see Figure \ref{FIGC12}.

The low-temperature limit of $c(T)$ can be obtained by a series expansion of $D(\varepsilon)$ at $\varepsilon=E_{\scriptsize min}$
of the form
\begin{equation}\label{lTser}
D(\varepsilon)= \sum_{\nu=0,1,2,\ldots}a_\nu\, \left(\varepsilon- E_{\scriptsize min}\right)^{\alpha\,\nu}
\;,
\end{equation}
where $\alpha=1$ or  $\alpha=1/2$.
At low temperatures, the Boltzmann factor effectively constrains the system to energies
only slightly above $E_{\scriptsize min}$, and thus truncation to a few terms of the series (\ref{lTser})
would yield the low-temperature limit of $c(T)$, and analogously for $c^\ast(T)$.
Considering only one term of (\ref{lTser}) yields the remarkably simple result
\begin{equation}\label{climT0}
  \lim_{T\rightarrow 0} c(T)=\alpha+1
  \;.
\end{equation}

For the example (\ref{choice}) we thus obtain
\begin{equation}\label{lTlim1}
  c(T)=2+\frac{4}{9} \left(2-\sqrt{2}\right) T+\frac{40}{27} \left(3-2 \sqrt{2}\right) T^2+O(T^3)
  \;,
\end{equation}
see Figure \ref{FIGC12}, left panel, where the limit $c(T=0)=2$ is due to (\ref{climT0}) and the fact that the Taylor series of
$D(\varepsilon)$ starts with a linear term at $\varepsilon= E_{\scriptsize min}$, i.~e., $\alpha=1$.

For the sign-inverted system the density of state function in the lowest energy interval will be of the form
\begin{equation}\label{dosinv}
  D^\ast(\varepsilon)=a_0 \sqrt{\varepsilon-E_{\scriptsize min}}
  \;,
\end{equation}
and hence the series (\ref{lTser}) will only consist of a single term with $\alpha =1/2$.
This yields the low-temperature ``flatfoot" limit
\begin{equation}\label{lTlim2}
 c^\ast(T)=\frac{3}{2}+\ldots
 \;,
\end{equation}
such that the correction terms are only due to the influence of higher energy intervals,
see Figure \ref{FIGC12}, right panel, and  Figure \ref{FIGC1}. The form of $c^\ast(T)$ is reminiscent
of the double peak Schottky-type structure of the specific heat known from other examples \cite{NJ02,SPSL16,KSM16}.
Furthermore, by means of Monte Carlo spin dynamics simulations we numerically calculate the specific heat $c(T)$ and analogously $c^\ast(T)$ according to
(\ref{defc}).

\section{Susceptibility}\label{sec:SU}

\begin{figure}[htb]
\centering
 \includegraphics[width=1.0 \linewidth ]{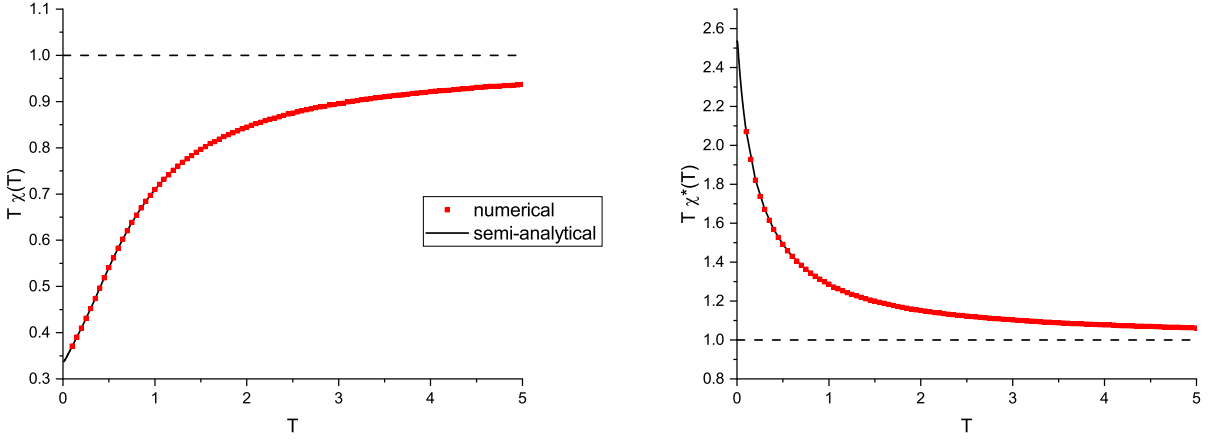}
 \caption{
Plot of the product of temperature and susceptibility  
$T\,\chi(T)$  for the choice of coupling constants (\ref{choice}) (left panel) as well as $T\,\chi^\ast(T)$
for the sign-inverted constants (right panel).
We observe a perfect agreement between the numerical values (red dots) and the semi-analytical ones (black curves).
The high temperature limit $\lim_{T\to\infty} T\,\chi(T)=1$ is indicated by black dashed lines. Also the limits $\lim_{T\to 0}T\,\chi(T)=1/3$ and
$\lim_{T\to 0}T\,\chi^\ast(T)=\frac{11}{6}+\frac{1}{\sqrt{2}}\approx 2.54044$ can be analytically calculated as explained in the text.
}
\label{FIGTC12}
\end{figure}

If we apply a magnetic field $B$ in $3$-direction, the Hamiltonian (\ref{defHam}) is modified by an additional Zeeman term to
\begin{equation}\label{defHB}
 H_B= H - {\mathbf S}^{(3)} B
 \;.
\end{equation}
The resulting magnetization $M(B)$ is given by
\begin{equation}\label{defmag}
 M(B)=\frac{1}{Z_B}\int_{\mathcal P}{\mathbf S}^{(3)}\,\exp\left( -\beta\,H_B\right)\,dV
 \;,
\end{equation}
with the modified partition function
\begin{equation}\label{defZB}
 Z_B=\int_{\mathcal P}\exp\left( -\beta\,H_B\right)\,dV
 \;.
\end{equation}
Due to the isotropy of the Hamiltonian (\ref{defHam}) the magnetization will vanish at $B=0$:
\begin{equation}\label{magzero}
  M(0)=0
  \;.
\end{equation}
Hence the first generally non-vanishing term of the Taylor expansion of $M(B)$ at $B=0$ will be the
``zero field susceptibility"
\begin{equation}\label{defchi}
  \chi(\beta)=\left.\frac{d M}{dB}\right|_{B=0}
  \;.
\end{equation}
After a short calculation using (\ref{magzero}) one obtains the well-known expression
\begin{equation}\label{chi1}
 \chi(\beta)= \frac{\beta}{Z_0}\int_{\mathcal P}{\mathbf S}^{(3)2}\,\exp\left( -\beta\,H\right)\,dV=:\beta\,\langle {\mathbf S}^{(3)2}\rangle
 \;.
\end{equation}
Again using isotropy of $H$ we have $\langle {\mathbf S}^{(1)2}\rangle=\langle {\mathbf S}^{(2)2}\rangle=\langle {\mathbf S}^{(3)2}\rangle$
and hence
\begin{equation}\label{chi2}
 \chi=\frac{1}{3}\beta\langle S^2\rangle\stackrel{(\ref{conserved2})}{=}\beta\,\langle 1+\frac{2}{3}\sigma\rangle
=
\beta\left(1+\frac{2}{3\,Z(\beta)}\int_\Sigma \sigma\,\exp\left(-\beta\, \varepsilon \right)
\,D(\sigma,\varepsilon)\,d\sigma\,d\varepsilon \right)
 \;,
\end{equation}
with $Z(\beta)$ given by (\ref{defZ}) and $\Sigma$ by (\ref{defSigma}). Hence also for the calculation of the zero field susceptibility
we can reduce the integration over the phase space to an integration over the two-dimensional domain $\Sigma$.

Although the values of $D(\sigma,\varepsilon)$ can be determined analytically, the double integral (\ref{chi2}) can only be performed
numerically. To this end we have constructed a grid covering
$\Sigma$ consisting of $86,159$ points. The double integral (\ref{chi2}) is then approximated by a sum over the grid.

The asymptotic behavior of the susceptibility for the limit values $T\to\infty$ and $T\to 0$
can be analytically determined. For $T\to \infty$ the leading term of (\ref{chi2})
is obtained by setting $\beta=0$ in the bracket. It follows that
\begin{equation}\label{racket0}
 \int_\Sigma \sigma\, D(\sigma,\varepsilon)\,d\sigma\,d\varepsilon\stackrel{(\ref{sigma0})}{=}0
 \;,
\end{equation}
and hence
\begin{equation}\label{limTinf}
  \lim_{T\to\infty}T\,\chi(T)=1
  \;,
\end{equation}
independent of the chosen coupling constants and in accordance with Curie's law.

For the limit $T\to 0$ we can assume that the density of states is more and more
concentrated in a region close to the ground state and therefore the variable $\sigma$ in (\ref{chi2})
will be asymptotically constant $\sigma\to\sigma_0$  and can be pulled out of the integral. This yields
\begin{equation}\label{limT0}
  \lim_{T\to 0}T\,\chi(T)=1+\frac{2}{3}\sigma_0
  \;.
\end{equation}
For the choice of coupling constants (\ref{choice}) the ground state is $\uparrow\downarrow\downarrow$ and hence $\sigma_0=-1$.
This yields  $\lim_{T\to 0}T\,\chi(T)=1-\frac{2}{3}=\frac{1}{3}$, see Figure \ref{FIGTC12}, left panel.
For the sign-inverted choice the ground state corresponds to the point ${\mathbf e}_4=(u_4,v_4,w_4)$ of the Gram set
according to (\ref{e4u} - \ref{e4w}) and hence $\sigma_0=u_4+v_4+w_4=\frac{1}{4} \left(5+3 \sqrt{2}\right)$.
This leads to $\lim_{T\to 0}T\,\chi^\ast(T)=\frac{11}{6}+\frac{1}{\sqrt{2}}\approx 2.54044$, see Figure \ref{FIGTC12}, right panel.
Furthermore, by means of Monte Carlo spin dynamics simulations we numerically calculate the product $T\chi(T)$ and analogously $T\chi^\ast(T)$ by using
(\ref{chi1}).

\section{Spin autocorrelation function}\label{sec:SA}

\begin{figure}[htb]
\centering
 \includegraphics[width=1.0 \linewidth ]{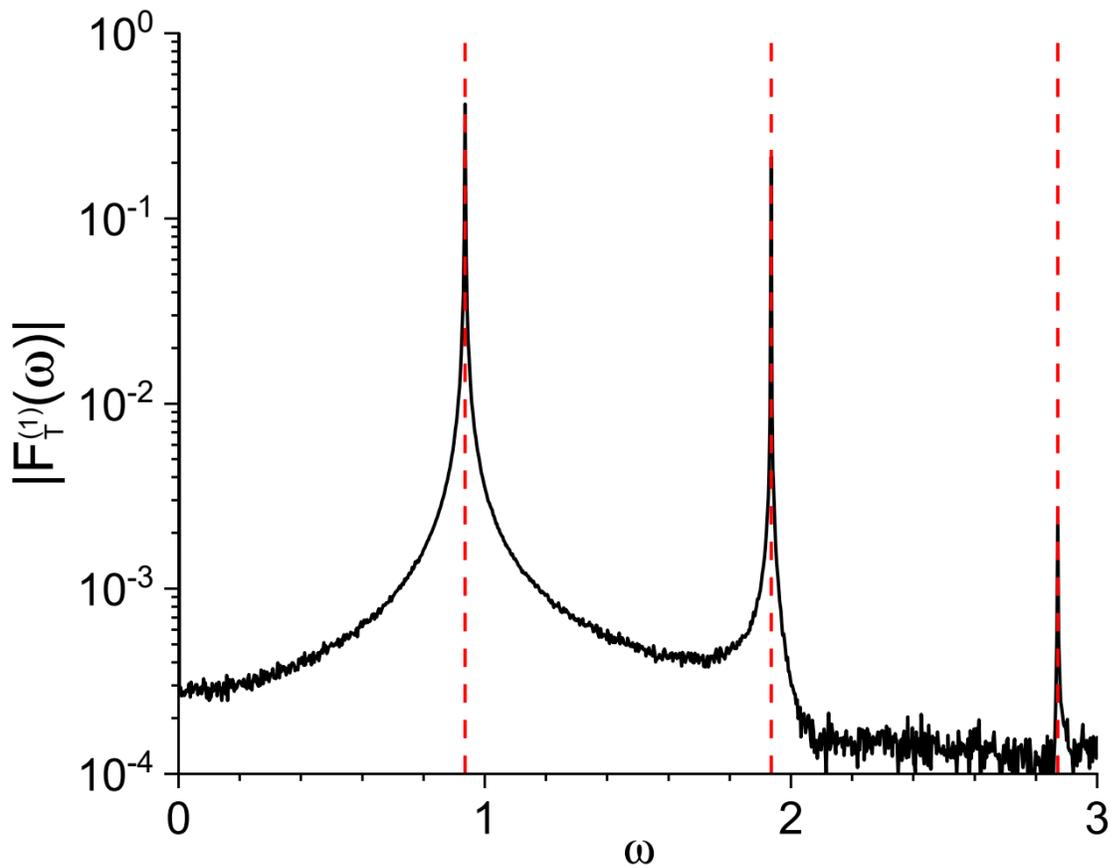}
 \caption{
Logarithmic plot of the power spectrum $\left| F^{(1)}_T(\omega) \right|$ of the spin triangle with coupling constants (\ref{choice})
and $T=10^{-4}$.
We observe three peaks at $\omega_1=0.9355, \,\omega_2=1.9355$ and $\omega_3=\omega_1+\omega_2=2.871$ (vertical dashed red lines). The third peak
is suppressed by thermal averaging and only visible due to the logarithmic scale.
}
\label{FIGACF1}
\end{figure}

The  autocorrelation function (acf) provides information about repeating patterns in the time evolution of a system.
In our case it will be defined as the thermal expectation value of the scalar product ${\mathbf s}_\mu(0)\cdot {\mathbf s}_\mu(t)$:
\begin{equation}\label{defauto}
 f^{(\mu)}_\beta(t)=\frac{1}{Z(\beta)} \int_{\mathcal P} {\mathbf s}_\mu(0)\cdot {\mathbf s}_\mu(t)\,\exp\left(-\beta H(s) \right)\,dV
 \;,
\end{equation}
for $\mu=1,2,3$.
Its absolute square value $\left|F^{(\mu)}_\beta(\omega)\right|^2$ of it Fourier transform $F^{(\mu)}_\beta(\omega)$ can be identified
with the ``spectral power density" by virtue of the Wiener-Khinchine theorem \cite{W30,K34}. We will also refer to
$\left|F^{(\mu)}_\beta(\omega)\right|$ as the ``power spectrum" without danger of confusion.

In order to calculate the thermal expectation value of the scalar product ${\mathbf s}_\mu(0)\cdot {\mathbf s}_\mu(t)$ numerically, we used the so-called ``Gibbs approach" \cite{LL99}, where the trajectories ${\mathbf s}_\mu(t)$ for the spins are calculated for the isolated system by solving the equations of motion (\ref{eom1} - \ref{eom3}) over a certain number of time steps numerically. The initial conditions for each trajectory are generated by a standard Monte Carlo simulation for
a temperature $T$. By averaging all generated trajectories over equivalent time intervals $t$ one obtains the canonical ensemble average $\langle {\mathbf s}_\mu(0)\cdot {\mathbf s}_\mu(t) \rangle$ at that temperature $T$. Although this procedure can easily be parallelized it is still computationally very much more demanding compared to the calculation of static thermal averages like the specific heat or the susceptibility. In addition to the generation of Monte Carlo samples the numerical time integration of the equations of motion for each sample needs to be done. However, since each point in time can be used as ${\mathbf s}_\mu(0)$ an additional averaging over equivalent time intervals within each trajectory can be exploited. This leads to an asymmetric quality improvement for $\langle {\mathbf s}_\mu(0)\cdot {\mathbf s}_\mu(t) \rangle$. With $N_t$ being the total number of points in time to be saved for each trajectory the calculation of $\langle {\mathbf s}_\mu(0)\cdot {\mathbf s}_\mu(t=1) \rangle$ is by a factor $N_t$ more accurate than $\langle {\mathbf s}_\mu(0)\cdot {\mathbf s}_\mu(t=N_t) \rangle$. This effect is especially notable for high temperatures where many Monte Carlo samples need to be generated in order to produce sufficient accurate thermal averages. In Figure \ref{FIGACF3} one can see that with increasing $t$ the oscillations become more noisy. In the frequency domain (see Figure \ref{FIGHTP}) we therefore find very accurate data for high frequencies for all temperatures whereas for low frequencies the data quality decreases with increasing temperatures.

\subsection{Short time autocorrelation}\label{sec:STA}

In this section we consider the
``short time autocorrelation function", that is, we restrict the domain of definition of the function
$t\mapsto f^{(\mu)}_\beta(t)$ to values comparable with the mean period ${\sf T}$ of the time evolution of the internal variables.

For small temperatures $T\to 0$ the power spectrum $\left|F^{(\mu)}_T(\omega)\right|$ will be markedly peaked at certain
``resonant frequencies"
$\omega_i$ that are typical for the time evolution close to the system's ground state.
At first glance, one might think to find $\omega_{\sf T}=\frac{2\pi}{\sf T}$,
evaluated in the ground state, as the resonant frequency,
but this is too simple thinking, since $\omega_{\sf T}$ is only the frequency of oscillation of the {\it internal} variables.
The global rotation in spin space, described by (\ref{eomext}), that corresponds to the external variables, also contributes to the
resonant frequencies.

In order to obtain closed formulas for the resonant frequencies in the limit $T\to 0$ we
will linearize the equations of motion (\ref{eom1} - \ref{eom3}) for solutions close to the ground state.
The details of the calculation depend on whether the ground state is collinear or coplanar.

\subsubsection{The case of collinear ground state}\label{sec:COL}
In this subsection we choose the coupling constants according to (\ref{choice}) and
thus obtain the ground state symbolized by $\downarrow\uparrow\uparrow$ with $E_{\scriptsize min}=-1-\sqrt{2}$.
We consider the time evolution of $s(t)$ for initial values close to the
ground state. To this end we write the three spin vectors
according to the ansatz
\begin{equation}\label{spinlin}
  {\mathbf s}_\mu(t)=\left(\begin{array}{c}
    \lambda\, s_\mu^{(1)}(t) \\
    \lambda\, s_\mu^{(2)}(t) \\
     h_\mu+\lambda^2\,s_\mu^{(3)}(t)
    \end{array} \right) +O\left(\lambda^2\right)
 \;,
\end{equation}
for $\mu=1,2,3,$ where the $h_\mu$ in the third component are chosen as $-1,1,1$
according to the ground state configuration $\downarrow \uparrow\uparrow$.
The terms of order $\lambda^2$ are only shown for the $3$rd component. These terms are obtained from the
$\lambda$-terms by solving the equations
\begin{equation}\label{linear}
  2\,h_\mu\,s_\mu^{(3)}(t) + \left(s_\mu^{(1)}(t)\right)^2 +\left(s_\mu^{(2)}(t)\right)^2 =0,\quad \mbox{for }\mu=1,2,3,
\end{equation}
that result from normalization of the ${\mathbf s}_\mu(t)$ up to $\lambda^2$-terms.
Obviously, the energy $\varepsilon$ of the configuration (\ref{spinlin})) satisfies $\varepsilon=E_{\scriptsize min}+O(\lambda^2)$.

We linearize the equations of motion (\ref{eom1} - \ref{eom3}) w.~r.~t.~$\lambda$ and obtain a differential equation of the form
\begin{equation}\label{lineom}
 \dot{\boldsymbol\xi}= A\,{\boldsymbol\xi}
 \;,
\end{equation}
where ${\boldsymbol\xi}$ is a vector comprising the six unknown terms of order $\lambda$ in (\ref{spinlin})
\begin{equation}\label{defxi}
{\boldsymbol\xi}(t)=\left(   s_{1}^{(1)}(t),\  s_{1}^{(2)}(t),  s_{2}^{(1)}(t),  s_{2}^{(2)}(t),  s_{3}^{(1)}(t),  s_{3}^{(2)}(t)\right)
\;,
\end{equation}
and $A$ is the real  matrix
\begin{equation}\label{defA}
 A=\frac{1}{2}\,\left(
\begin{array}{cccccc}
 0 & -1-2 \sqrt{2} & 0 & -\sqrt{2} & 0 & -1-\sqrt{2} \\
 1+2 \sqrt{2} & 0 & \sqrt{2} & 0 & 1+\sqrt{2} & 0 \\
 0 & \sqrt{2} & 0 & 1+\sqrt{2} & 0 & -1 \\
 -\sqrt{2} & 0 & -1-\sqrt{2} & 0 & 1 & 0 \\
 0 & 1+\sqrt{2} & 0 & -1 & 0 & 2+\sqrt{2} \\
 -1-\sqrt{2} & 0 & 1 & 0 & -2-\sqrt{2} & 0 \\
\end{array}
\right)
\;.
\end{equation}
$A$ has a double eigenvalue $0$ and two complex-conjugate pairs of imaginary eigenvalues of the form
\begin{equation}\label{eigA}
\pm {\sf i}\,\omega_1=\pm \frac{\sf i}{2} \left(-1+\sqrt{4+3 \sqrt{2}}\right),\quad
\pm {\sf i}\,\omega_2=\pm \frac{\sf i}{2} \left(1+\sqrt{4+3 \sqrt{2}}\right)
\;.
\end{equation}
This implies that the $1,2$-components of ${\mathbf s}_\mu$ will
perform superpositions of two harmonic oscillations about a mean value with the frequencies
$\omega_1=\frac{1}{2}\left(-1+\sqrt{4+3 \sqrt{2}}\right)=0.9355\ldots$
and
$\omega_2=\frac{1}{2}\left(1+\sqrt{4+3 \sqrt{2}}\right)=1.9355\ldots$.
We expand the autocorrelation ${\mathbf s}_\mu(0)\cdot {\mathbf s}_\mu(t)$ up to terms of order $\lambda^2$:
\begin{equation}\label{expandacf}
{\mathbf s}_\mu(0)\cdot {\mathbf s}_\mu(t)=
1+ \lambda^2\,\left(s_\mu^{(1)}(0)\,s_\mu^{(1)}(t) +s_\mu^{(2)}(0)\,s_\mu^{(2)}(t)+h_\mu \left(s_\mu^{(3)}(0)+s_\mu^{(3)}(t) \right)\right)
\;.
\end{equation}
From this expression one can read off the frequencies that will be possibly present in the autocorrelation function, namely
$\omega_1$ and $\omega_2$ due to the first two terms and the combinations $\omega_2\pm\omega_1$ due to $s_\mu^{(3)}(t)$.
Recall that the latter is a sum of $\left(s_\mu^{(i)}(t)\right)^2$-terms, $i=1,2,$ according to (\ref{linear}).

The two frequencies $\omega_1$ and $\omega_2$  correspond to two prominent peaks in Figure \ref{FIGACF1},
whereas the peak at  $\omega_1+\omega_2=\omega_{\sf T}=2.871\ldots$  seems to be suppressed by thermal averaging
and the resonant frequency $\omega_2-\omega_1=1$ is completely absent.
This will be explained in what follows.

First we may determine the six real solutions $s_\mu^{1,2}(t)$ by suitable linear combinations
of fundamental solutions  of (\ref{lineom})  using the eigenvectors of $A$. It turns out that
the coefficients of $\sin t$ and $\cos t$ in $(s_\mu^{1}(t))^2+(s_\mu^{2}(t))^2$ cancel and
hence  $s_\mu^{(3)}(t)$ will be of the form
\begin{equation}\label{smu3}
 s_\mu^{(3)}(t) =a_0 + a_1 \cos(\omega_{\sf T} t)+a_2 \sin(\omega_{\sf T} t)
 + a_3 \cos(\omega_1 t)+a_4 \sin(\omega_1 t)+ a_5 \cos(\omega_2 t)+a_6 \sin(\omega_2 t)
 \end{equation}
i.~e., without $\sin t$- and $\cos t$-terms. This already explains the absence of the resonant frequency $\omega=1$.

Next we consider the internal variables $u(t),\,v(t),\,w(t)$ resulting from the ansatz (\ref{spinlin})
and the corresponding standard configuration $r(t)$ according to (\ref{defr1} - \ref{defr3}). It is connected
to the original linearized solution  $(\ref{spinlin})$ by a time-dependent rotation $Z(t)$ such that $s(t)= Z(t)\,r(t)$, compare (\ref{solst}).
Due to the form of $Z(t)$, see (\ref{formZ}), the $3$rd row of $r(t)$ is left unchanged by $Z(t)$ and hence coincides with the
$3$rd row of $s(t)$. From the above considerations it follows that, up to terms linear  in $\gamma$,
 ${\mathbf s}_\mu^{(3)}(0)\,{\mathbf s}_\mu^{(3)}(t)={\mathbf r}_\mu^{(3)}(0)\,{\mathbf r}_\mu^{(3)}(t)$
will be of the form
\begin{equation}\label{comp3}
 {\mathbf r}_\mu^{(3)}(0)\,{\mathbf r}_\mu^{(3)}(t) =1+\lambda^2\left( a_0 + a_1 \cos(\omega_{\sf T} t)+a_2 \sin(\omega_{\sf T} t)
 + a_3 \cos(\omega_1 t)+a_4 \sin(\omega_1 t)+ a_5 \cos(\omega_2 t)+a_6 \sin(\omega_2 t)\right)
 \;.
\end{equation}

Let us consider the thermal average of (\ref{comp3}). According to Section \ref{sec:DS}
we may use the uniform distribution w.~r.~t.~the variables $\sigma, \varepsilon, \tau$ modulated by the Boltzmann factor
$\exp(-\beta\,\varepsilon)$. Here we have denoted the time coordinate by $\tau$ in order to distinguish it from the above
time evolution parameter $t$. We have to replace the initial value ${\mathbf r}_\mu^{(3)}(0)$
by ${\mathbf r}_\mu^{(3)}(\tau)$, and analogously ${\mathbf r}_\mu^{(3)}(t)$ by   ${\mathbf r}_\mu^{(3)}(\tau+t)$
where $\tau$ varies uniformly over the interval $0\le \tau\le {\sf T}$.
(\ref{comp3}) will then be replaced by
\begin{eqnarray}\nonumber
{\mathbf r}_\mu^{(3)}(\tau)\,{\mathbf r}_\mu^{(3)}(\tau+t)&=&
 1+\lambda^2\left( a_0 + a_1 \cos(\omega_{\sf T}(t+\tau))+a_2 \sin(\omega_{\sf T} (t+\tau))
 + a_3 \cos(\omega_1 (t+\tau))+a_4 \sin(\omega_1 (t+\tau))\right.\\
 \label{acf4}
&& \left. + a_5 \cos(\omega_2(t+\tau))+a_6 \sin(\omega_2 (t+\tau))\right)
 \;.
\end{eqnarray}
The thermal average
is achieved by integrating (\ref{acf4}) over $0\le \tau\le {\sf T}$
(followed by integrations over $\sigma$ and $\varepsilon$ and multiplication with a suitable normalization factor).
We conclude that the $\tau$-integral of (\ref{acf4}) will be of the form
\begin{equation}\label{intacf}
 \frac{1}{\sf T}\int_{0}^{{\sf T}}{\mathbf r}_\mu^{(3)}(\tau)\,{\mathbf r}_\mu^{(3)}(\tau+t)\,d\tau
 = A_0+\sum_{i=1}^{2} A_i \cos(\omega_i t)+ B_i \sin(\omega_i t)
\;.
\end{equation}
Here we have used, first, that the mean values of $\sin \omega_{\sf T}\tau $ and $\sin \omega_{\sf T}(\tau+t) $ over a full period ${\sf }$ vanish.
Second, for frequencies $\omega\neq \omega_{\sf T}$ the following holds:
\begin{equation}\label{meaneq}
 \frac{1}{\sf T}{\int_0^{\sf T} \sin (\omega  (t+\tau )) \, d\tau } =
 \frac{(1-\cos (\omega  {\sf T})) \cos (\omega  t)+\sin (\omega {\sf T}) \sin (\omega t)}{\omega {\sf T}}
 \;.
\end{equation}

This shows that the thermal average of ${\mathbf r}_\mu^{(3)}(0)\,{\mathbf s}_\mu^{(3)}(t)$ reduces to a constant contribution
plus terms that oscillate with the resonant frequencies $\omega_1$ and $\omega_2$, but
the resonant frequency $\omega_{\sf T}$ will be suppressed. This can be seen in Figure \ref{FIGACF1} where the thermal average
has been calculated numerically and a tiny peak at $\omega=\omega_{\sf T}=2.871\ldots$ is still visible.

\subsubsection{The case of coplanar ground state}\label{sec:COP}

\begin{figure}[htb]
\centering
 \includegraphics[width=0.7 \linewidth ]{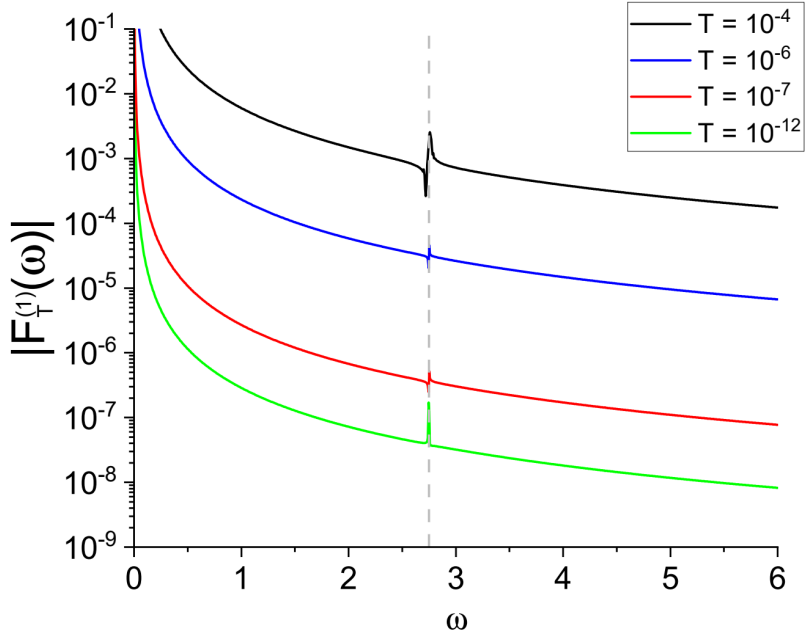}
 \caption{
 Logarithmic plot of the spectral power density $\left| F^{(1)}_T(\omega) \right|$ of the spin triangle
with sign-inverted coupling constants (\ref{choice})
and different low temperatures (see inset).
We observe a prominent peak at $\omega_1\approx 2.749 $ (vertical dashed line). Another posible peak at $2\omega_1\approx 5.498$
is suppressed by thermal averaging and not visible.
}
\label{FIGACF2}
\end{figure}

We consider the example of sign-inverted coupling constants (\ref{choice}). The coplanar ground state
$g=\left({\mathbf g}_1, {\mathbf g}_2, {\mathbf g}_3,\right)$ may
be chosen as
\begin{equation}\label{copground}
  g=\left(
\begin{array}{ccc}
 2-\frac{3}{2 \sqrt{2}} & \frac{1}{4} \left(3 \sqrt{2}-2\right) & 1 \\
 \frac{1}{2} \sqrt{12 \sqrt{2}-\frac{33}{2}} & \sqrt{\frac{3}{2
   \sqrt{2}}-\frac{3}{8}} & 0 \\
 0 & 0 & 0 \\
\end{array}
\right)
\;,
\end{equation}
satisfying $\varepsilon=E_{\scriptsize min}=\frac{1}{8} \left(-10-\sqrt{2}\right)=-1.42678\ldots$
and $\sigma=\frac{1}{4} \left(5+3 \sqrt{2}\right)=2.31066\ldots$. We choose orthogonal unit vectors
${\mathbf h}_\mu\equiv (0,0,1)^\top$
and ${\mathbf k}_\mu={\mathbf g}_\mu\times {\mathbf h}_\mu$ that span the tangent plane
of the unit sphere at ${\mathbf g}_\mu$ for $\mu=1,2,3$ and,
similarly as for the collinear ground state, define a spin configuration $s(t)$ close to the ground state by
\begin{equation}\label{defst}
 {\mathbf s}_\mu(t)=\left( 1+\lambda^2 s_{\mu}^{(3)}(t) \right){\mathbf g}_\mu
 +\lambda\left(  s_{\mu}^{(1)}(t){\mathbf h}_\mu+ s_{\mu}^{(2)}(t){\mathbf k}_\mu\right)
 +O(\lambda^2)
 \;,
\end{equation}
for $\mu=1,2,3$. Normalization of ${\mathbf s}_\mu(t)$ up to the order of $\lambda^2$ implies
\begin{equation}\label{smu3}
  s_{\mu}^{(3)}(t)=-\frac{1}{2}\left( \left( s_{\mu}^{(1)}(t) \right)^2+\left(s_{\mu}^{(2)}(t) \right)^2\right)
\;,
\end{equation}
for $\mu=1,2,3$. Linearization of the equations of motion (\ref{eom1} - \ref{eom3}) yields
\begin{equation}\label{lincop}
  \dot{\boldsymbol\xi}=B\,{\boldsymbol\xi}
  \;,
\end{equation}
where
${\boldsymbol\xi}$ is defined as in  (\ref{defxi})
and
\begin{equation}\label{defB}
B=\frac{1}{2}\,
\left(
\begin{array}{cccccc}
 0 & 2+\sqrt{2} & 0 & \frac{1}{4} \left(\sqrt{2}-6\right) & 0 & \frac{1}{4}
   \left(-2-5 \sqrt{2}\right) \\
 -2-\sqrt{2} & 0 & \sqrt{2} & 0 & 1+\sqrt{2} & 0 \\
 0 & \frac{1}{4} \left(\sqrt{2}-6\right) & 0 & 2-\sqrt{2} & 0 & \frac{1}{4}
   \left(3 \sqrt{2}-2\right) \\
 \sqrt{2} & 0 & \sqrt{2}-2 & 0 & -1 & 0 \\
 0 & \frac{1}{4} \left(-2-5 \sqrt{2}\right) & 0 & \frac{1}{4} \left(3
   \sqrt{2}-2\right) & 0 & 1+\frac{1}{\sqrt{2}} \\
 1+\sqrt{2} & 0 & -1 & 0 & -1-\frac{1}{\sqrt{2}} & 0 \\
\end{array}
\right)
\;.
\end{equation}
The eigenvalues of $B$ are $0$ (with fourfold algebraic degeneracy) and
\begin{equation}\label{eigB}
  \pm{\sf i}\,\omega_1 = \pm \frac{i}{2} \sqrt{\frac{35}{2}+9 \sqrt{2}}
  \;.
\end{equation}
This yields a harmonic oscillation of $s(t)$ with frequency $\omega_1\approx 2.749$ which corresponds
to the resonant frequency in the power spectrum, see Figure \ref{FIGACF2}, where the double frequency
$2\,\omega_1\approx 5.498$ is thermally suppressed.

\subsection{Long time autocorrelation}\label{sec:LTA}

\subsubsection{Long time limit}\label{sec:LTL}

\begin{figure}[htb]
\centering
 \includegraphics[width=0.95 \linewidth ]{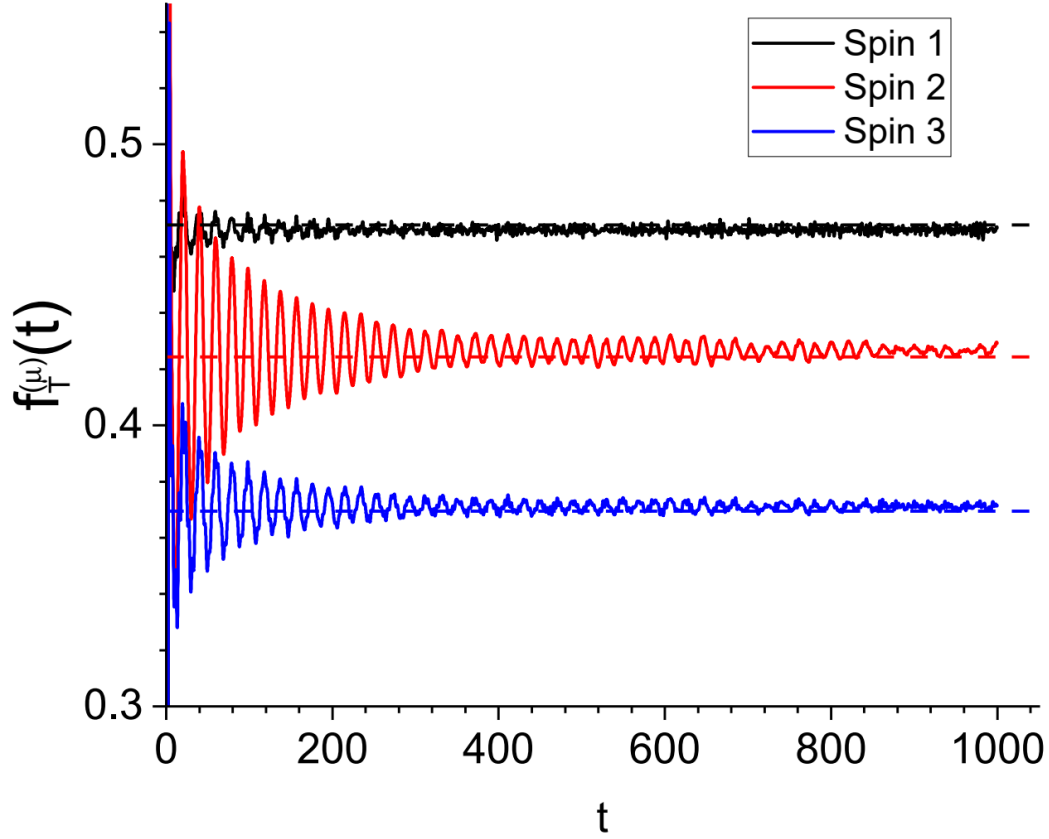}
 \caption{
 Plot of the autocorrelation functions $f_T^{(\mu)}(t),\,\mu=1,2,3,$ for the standard example (\ref{choice})
and temperature $T=10^6$ together with the asymptotic values $f^{(1)}\approx 0.47138,\,f^{(2)}\approx  0.424294$ and
$f^{(3)}\approx 0.369464$
calculated semi-analytically.
}
\label{FIGACF3}
\end{figure}

For long times, $t\to\infty$, the autocorrelation function assumes constant values. In order to calculate these values
we adopt the ``Floquet form" of the time dependence of the spin configuration
\begin{equation}\label{Floquet}
  s(t) = {\mathcal R}({\mathbf S},\alpha t)\,\widetilde{r}(t)
  \;,
\end{equation}
where ${\mathcal R}({\mathbf S},\alpha t)$ denotes a rotation about the total spin axis ${\mathbf S}$
with constant angular velocity $\alpha$ and $\widetilde{r}(t)$ will be ${\sf T}$-periodic, see \cite{S21}.
Without loss of generality we may choose the initial
configuration $\widetilde{r}(0)$ of the form (\ref{defr1} - \ref{defr3}) which entails ${\mathbf S}=(0,0,S)^\top$
and further
\begin{equation}\label{formofR}
  {\mathcal R}({\mathbf S},\alpha t)=
  \left(
  \begin{array}{ccc}
    \cos\alpha t & -\sin\alpha t & 0 \\
   \sin\alpha t & \cos\alpha t & 0 \\
    0 & 0 & 1
  \end{array}
  \right)
  \;.
\end{equation}
For the ${\sf T}$-periodic components of $\widetilde{r}(t)$ we use the Fourier series representation
\begin{equation}\label{Fourier}
  \widetilde{r}_{i\mu} = \sum_{n\in{\mathbbm Z}} a_{i\mu n}\,\exp\left( {\sf i}\,\omega \, n\, t\right)
  \;,
\end{equation}
with $\omega\equiv 2\pi/{\sf T}$ and
\begin{equation}\label{cc}
a_{i,\mu,-n}=\overline{a_{i,\mu,n}}
\end{equation}
for all $1\le i,\mu\le 3$ and $n\in {\mathbbm Z}$ due to $\widetilde{r}_{i\mu} \in{\mathbbm R}$.
Recall that the autocorrelation function is the thermal average of
\begin{equation}\label{scalarprod}
 {\mathbf s}_\mu(0)\cdot {\mathbf s}_\mu(t) =\sum_{i=1}^3 s_{i\mu}(0)\, s_{i\mu}(t)=:\sum_{i=1}^{3}sp_i^{(\mu)}
 \;.
\end{equation}
The third term $sp_3^{(\mu)}$ of this sum is particularly simple since it is independent of $\alpha$ due to (\ref{formofR}).
It reads
\begin{equation}\label{thirdterm}
 sp_3^{(\mu)}=s_{3\mu}(0)\,s_{3\mu}(t)= \widetilde{r}_{3\mu}(0)\,\widetilde{r}_{3\mu}(t)\stackrel{(\ref{Fourier})}{=}
 \sum_{nm}a_{3\mu n}\,a_{3\mu m}\,\exp\left( {\sf i}\,\omega \, n\, t\right)
 \;.
\end{equation}
As the first step in calculating the thermal average we consider the $\tau$-translate of this term over one period $0\le \tau\le {\sf T}$
\begin{equation}\label{timetrans}
 s_{3\mu}(\tau)\,s_{3\mu}(t+\tau)=\sum_{nm}a_{3\mu n}\,a_{3\mu m}\,\exp\left({\sf i}\,\omega \, m\,\tau+ {\sf i}\,\omega \, n\, (t+\tau)\right)
 \;,
\end{equation}
and integrate it with the result
\begin{equation}\label{integrate}
\left\langle sp_3^{(\mu)} \right\rangle_{\sf T}:=\frac{1}{\sf T}\int_{0}^{\sf T} s_{3\mu}(\tau)\,s_{3\mu}(t+\tau)\,d\tau = \sum_n a_{3\mu n}\,\overline{a_{3\mu n}}
\,\exp\left({\sf i}\,n\,\omega\,t \right)
\;,
\end{equation}
using (\ref{cc}). The $n=0$ term of this series does not depend on $t$ and $\omega$. If we split off this term from the autocorrelation function
it is plausible that the Fourier transform of the remainder will be an $L^1$-integrable function  of $\omega$ and hence
goes to zero for $t\to\infty$  (Riemann-Lebesgue lemma). This leads to
\begin{equation}\label{limtinf}
 \lim_{t\to\infty}  f^{(\mu)}_\beta(t)= \left\langle \left|a_{3\mu 0}\right|^2 \right\rangle_\beta
 \;,
\end{equation}
where $ \left\langle \ldots \right\rangle_\beta$ denotes the thermal average obtained by further integrations over $\varepsilon$ and $\sigma$.

For later purpose we note the analogous result for the sum of the other two terms $sp_1^{(\mu)}+sp_2^{(\mu)}$:
\begin{eqnarray}\nonumber
 \left\langle sp_1^{(\mu)} +sp_2^{(\mu)}\right\rangle_{\sf T}&=&
 \frac{1}{4}\sum_n\left[ \left(\left| a_{1\mu n}+{\sf i}\,a_{2\mu n}\right|^2+ \left| a_{2\mu n}-{\sf i}\,a_{1\mu n}\right|^2\right)
 \exp\left({\sf i}\,(n\omega +\alpha)t \right)\right.\\
 \label{sp12}
 &&\left. +
  \left(\left| a_{1\mu n}-{\sf i}\,a_{2\mu n}\right|^2+ \left| a_{2\mu n}+{\sf i}\,a_{1\mu n}\right|^2\right)
 \exp\left({\sf i}\,(n\omega -\alpha)t \right)
 \right]
 \;.
\end{eqnarray}
The $n=0$ term of this series can be written as
\begin{equation}\label{sp12n0}
 \left(\left| a_{1\mu 0}\right|^2+\left| a_{2\mu 0}\right|^2\right) \cos \alpha t
 \;,
\end{equation}
using that $a_{i\mu 0}$ is real due to (\ref{cc}).\\

\begin{figure}[htb]
\centering
 \includegraphics[width=0.95 \linewidth ]{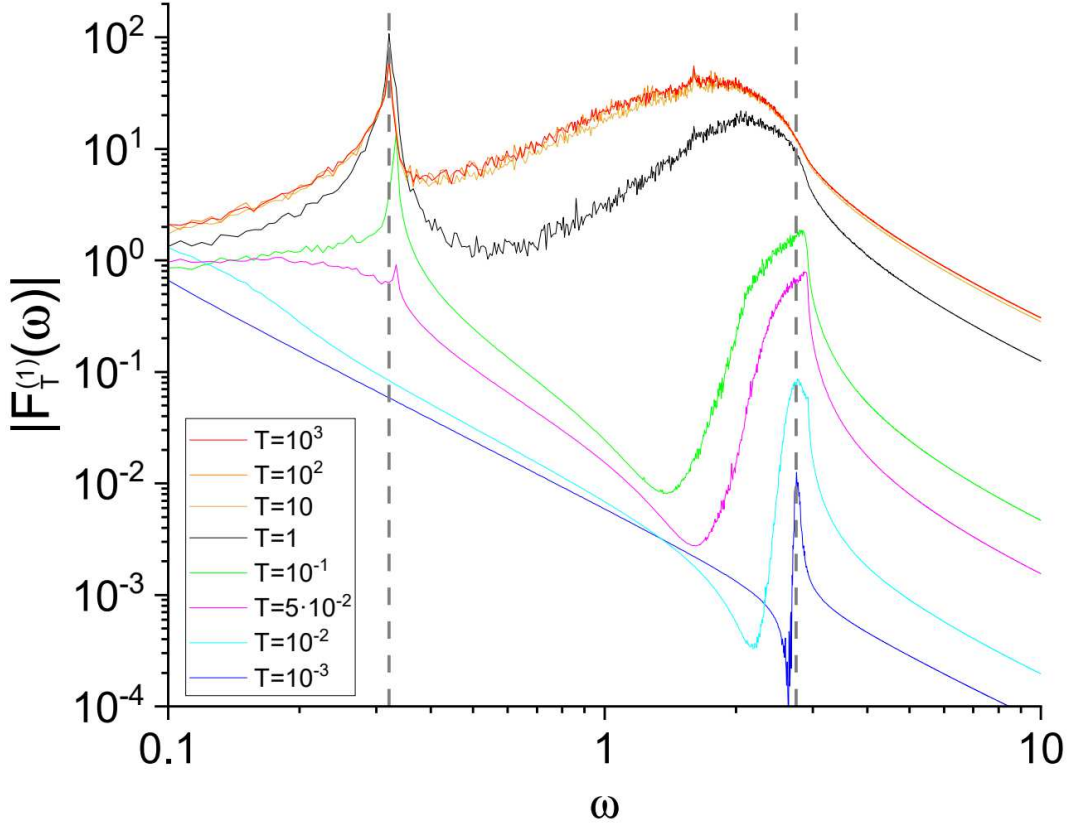}
 \caption{
 Plot of the power spectrum $|F_T^{(1)}(\omega)|$ for the sign-inverted standard example (\ref{choice})
 in double logarithmic scale for different temperatures (see inset). We observe that
 the low temperature peak at $\omega\approx 2.749$ due to the spin wave excitation, see also Figure \ref{FIGACF2},
 broadens for increasing temperature and a high temperature peak
 at $\omega_{HT}\approx 0.321$ arises (vertical dashed lines).
}
\label{FIGHTP}
\end{figure}

\begin{figure}[htb]
\centering
 \includegraphics[width=0.95 \linewidth ]{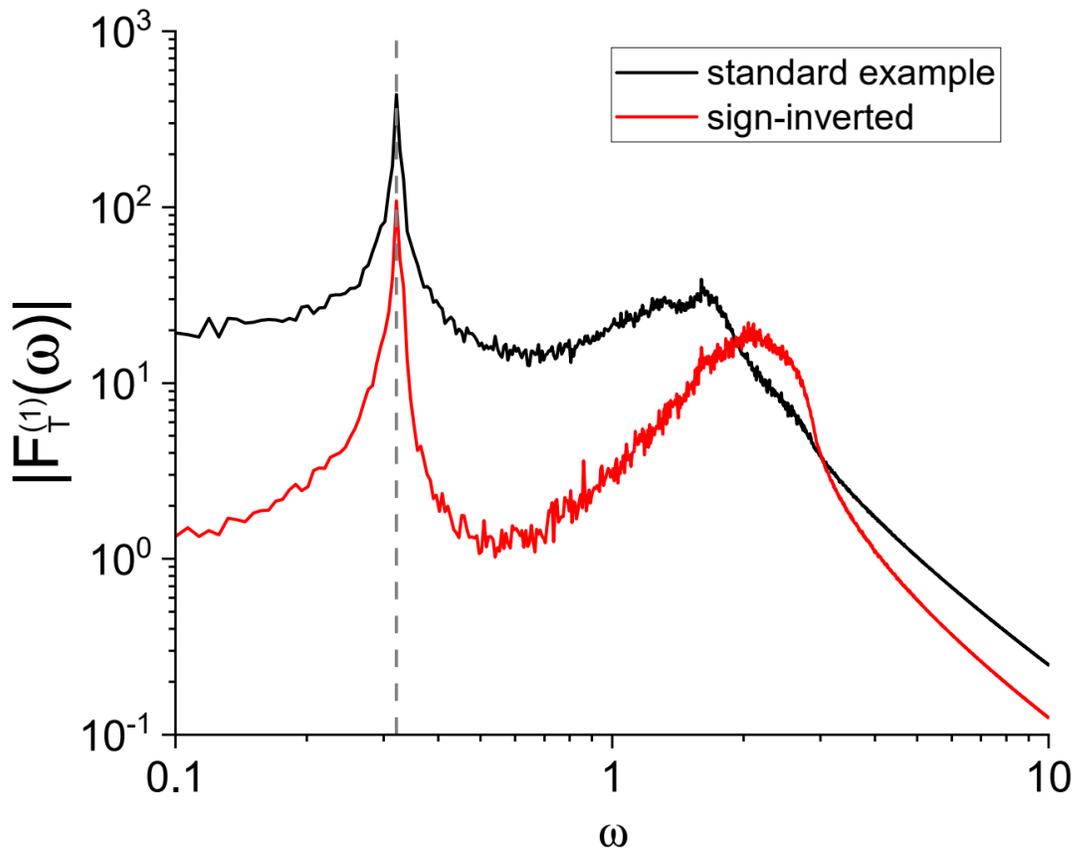}
 \caption{
 Doubly logarithmic plot of the power spectrum $|F_T(\omega)|$ for the standard example (\ref{choice}) and the sign-inverted example
 for the temperature $T=1$. We observe high temperature peaks for both systems
 at $\omega_{HT}\approx 0.321$  (vertical dashed line), see also Figure \ref{FIGHTP}.
}
\label{FIGACT2}
\end{figure}

The Fourier coefficient $a_{3\mu 0}$ is the time average of $\widetilde{r}_{3\mu}(t)$ over one period ${\sf T}$.
Since $\widetilde{r}_{3\mu}(t)$ is an affine function
of $x(t)$, see (\ref{defx}) and (\ref{defr1} - \ref{defr3}), this time average can be explicitly determined by means of
\begin{equation}\label{timeaverage}
\frac{1}{\sf T}\int_0^{\sf T}x(t)\,dt \stackrel{(\ref{solxt})}{=} \frac{1}{\sf T}\int_0^{\sf T} \wp \left(t+t_0;g_2,g_3\right)\,dt
=-\frac{2}{\sf T} \zeta \left(\frac{\sf T}{2};g_2,g_3\right)
\;,
\end{equation}
where the last identity follows from \cite[23.14.1]{NIST21}
and  $\zeta \left(\frac{\sf T}{2};g_2,g_3\right)$ denotes the {\em Weierstrass zeta function}.

It follows that the long time limit (\ref{limtinf}) can be semi-analytically calculated by a numerical
integration  over $\varepsilon$ and $\sigma$ using the density of states function $D(\sigma,\varepsilon)$ according to
(\ref{Dse}). The results for the standard example (\ref{choice}) and the high temperature limit $T\to \infty$
together with the numerical calculation of  $f^{(\mu)}_\beta(t)$ for $\mu=1,2,3$ and $T=1/\beta=10^6$ are shown in Figure \ref{FIGACF3}.

\subsubsection{High temperature peak of the spectral power density}\label{sec:HTP}

\begin{figure}[htb]
\centering
 \includegraphics[width=0.7 \linewidth ]{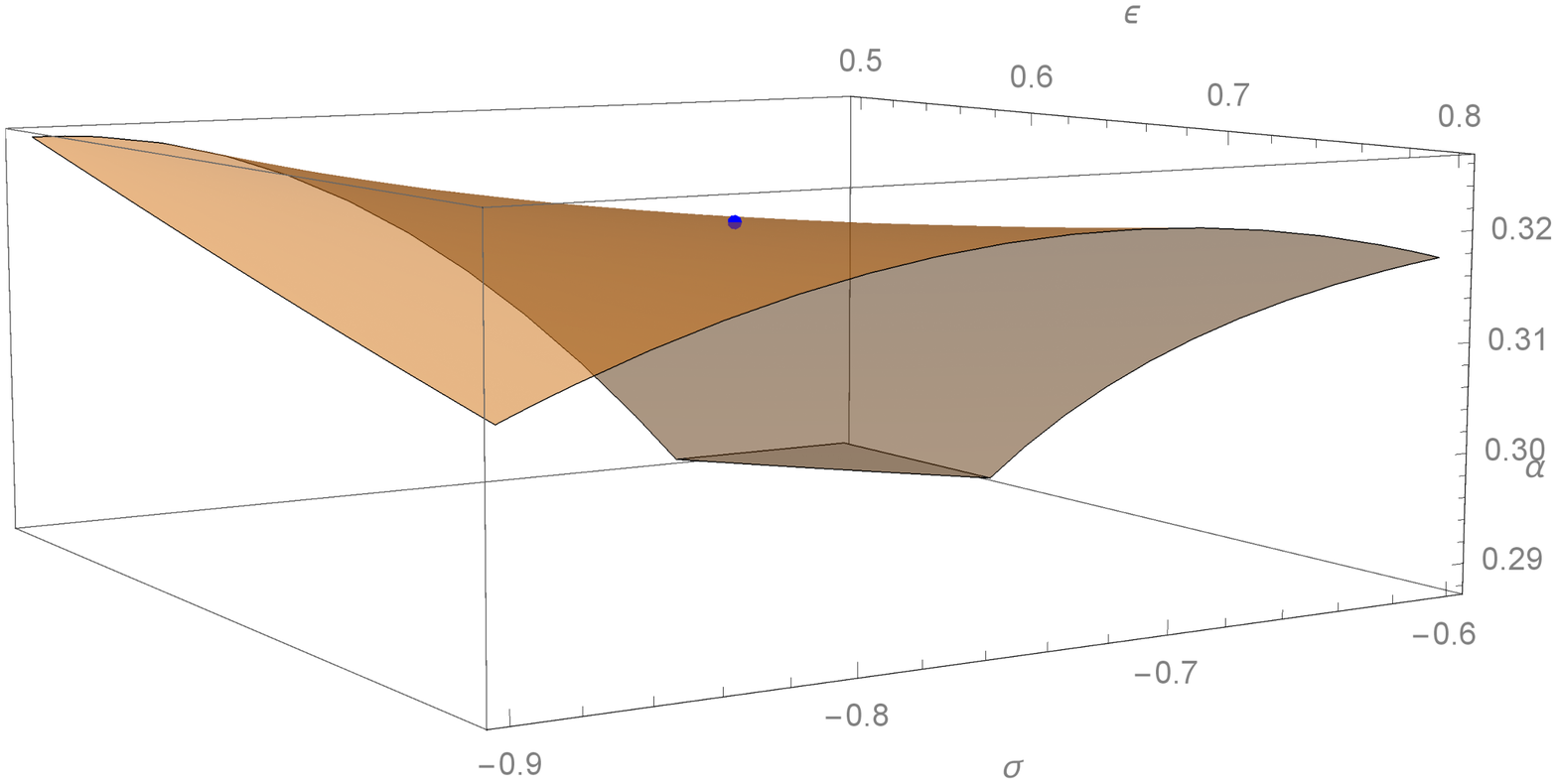}
 \caption{
 Plot of the graph of the angular velocity $\alpha(\varepsilon,\sigma)$ according  to (\ref{Floquet}) and (\ref{formofR})
 for the standard example (\ref{choice}) together with a saddle point at
 $\left(\varepsilon_0=0.707122,\,\sigma_0= -0.774964,\,\alpha_0=0.321055\right)$ (blue point).
}
\label{FIGSP}
\end{figure}

\begin{figure}[htb]
\centering
 \includegraphics[width=0.7 \linewidth ]{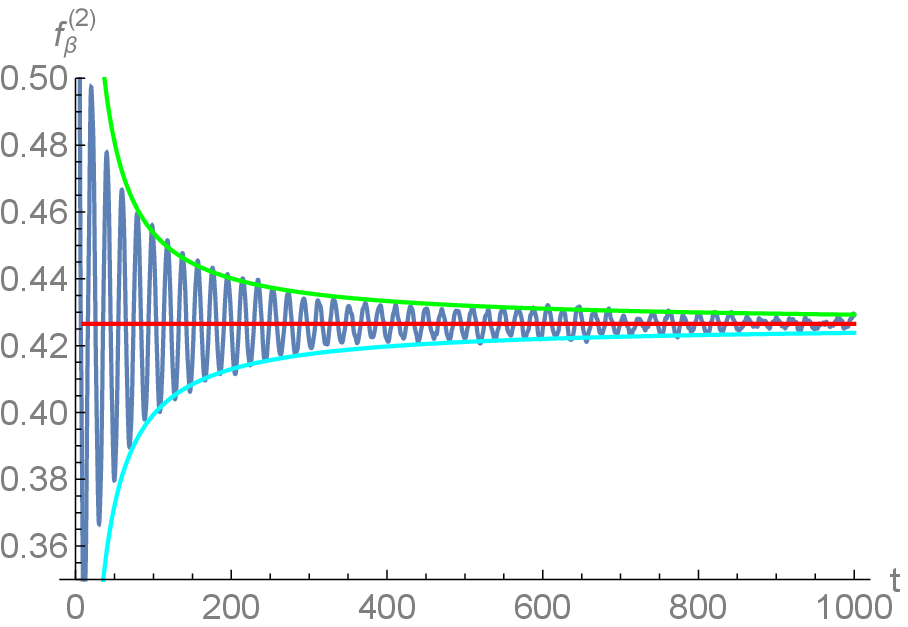}
 \caption{
 Plot of the autocorrelation function $f^{(2)}_T(t)$ for the standard example (\ref{choice})
 numerically calculated for the temperature $T=10^{6}$, see also Figure \ref{FIGACF3}.
 We observe a damped harmonic oscillation about the asymptotic value $f^{(2)}_T(\infty)=0.42654$ (red line) that decays proportional to the power $1/t$,
 see the (green and cyan) envelopes. The value of $f^{(2)}_T(\infty)$ has been determined numerically and approximately agrees
 with the semi-analytical result $f^{(2)}\approx  0.424294$ shown in Figure  \ref{FIGACF3}.
}
\label{FIGACF4}
\end{figure}

Numerical examples show that, for a certain domain of coupling constants,
the long time limit of the acf is assumed in the form of a damped harmonic oscillation, see Figure \ref{FIGACF3}.
The frequency of this oscillation gives rise to a peak at $\omega=\omega_{HT}$ in the spectral power density that is most clearly visible
for high temperatures (HT), see Figures \ref{FIGHTP} and \ref{FIGACT2}.
Typically, $\omega_{HT}$ is smaller than the frequency of the oscillation of the internal variables
$\omega_{\sf T}$ by a factor $10$ or so. Its origin lies in the time evolution of the external variables, or,
more precisely, in the mean rotation frequency $\alpha$ according to (\ref{Floquet}) and (\ref{formofR}).

We conjecture that, analogous to the long time limit of the acf, the damped harmonic oscillation regime can be described by
considering the $n=0$ terms (\ref{sp12n0}) of the series (\ref{sp12}) followed by thermal averaging:
\begin{equation}\label{dor1}
  f^{(\mu)}_\beta(t)\simeq  \left\langle \left|a_{3\mu 0}\right|^2 \right\rangle_\beta
  + \left\langle  \left(\left| a_{1\mu 0}\right|^2+\left| a_{2\mu 0}\right|^2\right) \cos \alpha t\right\rangle_\beta
  ,\quad t\to\infty
  \;.
\end{equation}
In the second term of this expression, $\alpha$ as well as $\left| a_{1\mu 0}\right|^2+\left| a_{2\mu 0}\right|^2$ are
functions on the domain $\Sigma$ of the values $(\varepsilon,\sigma)$
of the conserved quantities, see (\ref{defSigma}). We may interpret this term as the expectation value of the random variable
$\cos\, \alpha\,t$
w.~r.~t.~the (not normalized) probability distribution
\begin{equation}\label{probdist}
\widetilde{\rho}( \varepsilon,\sigma ):= \left(\left| a_{1\mu 0}\right|^2+\left| a_{2\mu 0}\right|^2\right)\,\exp(-\beta\,\varepsilon)\,D(\varepsilon,\sigma)
\end{equation}
defined on  $\Sigma$.

Numerical investigations show that in the cases where the mentioned effect occurs the distribution of the values of $\alpha$ has a sharp peak at $\omega=\omega_{HT}$.
This would explain that the dominant contribution to (\ref{dor1}) will be a damped harmonic oscillation with frequency $\omega_{HT}$.
Looking for a deeper explanation we found that the peak can be attributed to a
{\em saddle point} of the function $\alpha(\varepsilon,\sigma)$ at $(\varepsilon_0,\sigma_0)$.
Mathematically, any random variable $f$ of two arguments $x,y$ with a saddle point at $(x_0,y_0)$
has a distribution with a logarithmic singularity at $u_0=f(x_0,y_0)$, see Appendix \ref{sec:SP}.
Numerically, the singularity of the distribution shows up as a peak value.
In our case we may set $f(x,y)=\alpha(\varepsilon,\sigma)$ and determine the saddle point configurations numerically.
In the standard example of coupling constants (\ref{choice}) this is any spin configuration compatible with
the values $\varepsilon_0=0.707122$, $\sigma_0= -0.774964$  and consequently $\alpha_0=0.321055$, see Figure \ref{FIGSP},
in accordance with the position of the HT peak shown in Figure \ref{FIGACT2}.

It remains to be investigated what the decay of the acf looks like for $t\to \infty$.
We expect an algebraic decay since we are dealing with a completely integrable system.
Numerical evidence points to an algebraic decay of the type $1/t$.
To investigate this question further, we consider the r.~h.~s.~of (\ref{dor1}) and write the $t$-dependent part as
(the real part) of the Fourier transform $\widehat{\rho}(t)$  of a function $\rho(\omega)$ with a logarithmic singularity at $\omega=\omega_{HT}$.
This means that $\rho(\omega)$ is of the form $\rho(\omega)=\phi(\omega)\,\log|\omega-\omega_{HT}|$,
where $\phi(\omega)$ is sufficiently smooth.
Then it can be shown that the asymptotic form of $\widehat{\rho}(t)$ is proportional to $\frac{1}{t}\,\cos \omega_{HT} t$,
see (\ref{Erd6}), independent of the spin number $\mu$ and the temperature $T=1/\beta$.
This is consistent with the numerical results shown in Figure \ref{FIGACF4} for the standard example, $T=10^6$, and the second spin.
For the other two spins the results are less informative due to larger statistical fluctuations.

If our explanation of the HT peak by means of a saddle point of $\alpha(\varepsilon,\sigma)$ is correct,
we would expect that the peak does not appear for all values of the coupling constants but only for a certain domain.
It may happen that the saddle point moves to the boundary of $\Sigma$ and then disappears for certain values of the $J_i$.
In fact this has been observed. For the special case of $J_1=-1/2$ and $J_2=1$ we find a sharp HT peak only for $0.3 \lesssim J_3 \lesssim 0.7875$
and $J_3\gtrsim 1.35$.
The exact extent of the HT peak phase and the occurrence of other phases are not investigated here due to space limitations.

\section{Summary}\label{sec:SUM}
Although not typical, integrable systems are an interesting topic since
dynamical and thermodynamic quantities can be exactly calculated either analytically or semi-analytically in the form of integrals.
In the present case of a classical Heisenberg spin triangle with coupling constants $J_1,J_2,J_3$ we have recapitulated the recently obtained solution
of the equation of motion in terms of Weierstrass elliptic functions for the internal variables and certain integrals for the external variables.
For the density of states, specific heat and zero field susceptibility only the internal variables are needed. We could not provide
closed formulas that hold for all $J_1,J_2,J_3$ but have to proceed from case to case. We illustrate this procedure mainly for a standard example with a collinear ground state and the sign-reversed example with a coplanar one. The results are consistent with numerical Monte Carlo simulations.

The external variables come into play when the autocorrelation function (acf) of the system is examined. There are two cases that need to be investigated separately:
\begin{enumerate}
  \item The short time acf for low temperatures that shows peaks at frequencies that can be calculated from a linearized equation of motion, and
  \item the long time acf for high temperatures that, for certain values of the coupling constants, shows a damped harmonic oscillation about an asymptotic limit.
\end{enumerate}
In the latter case, we observed a high-temperature (HT) peak of the power spectrum that can be explained by a saddle point of the mean rotational frequency $\alpha(\varepsilon,\sigma)$ of the spin triangle as a function of the constants of motion $\varepsilon,\sigma$. Such a saddle point leads to a logarithmic singularity of the power spectrum and an algebraic $1/t$-decay of the acf according to a theorem of Erd\'{e}lyi, see appendix \ref{sec:ALS}.
A complete study of the HT phase and other phases of acf would be the subject of future work.

\appendix

\section{Details of the explicit time evolution}\label{sec:V}

Instead of $u$ we will use the variable $x$ given by
\begin{equation}\label{defx}
x=x_0+g\,u
 \;,
\end{equation}
where the constants $x_0$ and $g$ will be determined later such that the
\textit{Weierstrass differential equation} \cite[23.3.10]{NIST21} is obtained.
Also $v$ and $w$ can be linearly expressed in terms of the variable $x$ in the form:
\begin{eqnarray}\label{x2v}
 v&=& \frac{J_3-J_1}{J_2-J_3}\frac{x}{g}+v_0,\\
 \label{x2w}
 w&=& \frac{J_1-J_2}{J_2-J_3}\frac{x}{g}+w_0
 \;.
\end{eqnarray}
We consider the time derivative of $x$:
\begin{eqnarray}
\label{xd1}
 \dot{x} &\stackrel{(\ref{defx})}{=}& g\,\dot{u}\\
 \label{xd2}
   &\stackrel{(\ref{udot})}{=}&\pm g\,\left( J_3-J_2\right)\, \sqrt{ 1-u^2-v^2-w^2+2 u v w}\stackrel{(\ref{defdelta})}{=}g\,\left( J_3-J_2\right)\,\delta
  \;.
\end{eqnarray}

By substituting (\ref{x2v}), (\ref{x2w}) and (\ref{defx}), the square of (\ref{xd2})
can be written as a $3^{rd}$ order polynomial $\Pi(x)$.
$g$ and $x_0$ will be chosen such that the cubic term
of $\Pi(x)$ reads $4\,x^3$ and the quadratic
term of $\Pi(x)$  vanishes and hence
\begin{equation}\label{sdPi}
\left( \frac{dx}{dt}\right)^2=  \dot{x}^2=g^2\,\left( J_3-J_2\right)^2\,\delta^2 =\Pi(x) = 4 x^3-g_2 x-g_3
\;.
\end{equation}
This is achieved by setting
\begin{equation}\label{solg}
  g=-\frac{1}{2} \left(J_1-J_2\right) \left(J_1-J_3\right)
  \;,
\end{equation}
and
\begin{equation}\label{solx0}
  x_0=\frac{1}{6} \left(-J_1^2-J_2^2-J_3^2+J_1 J_3+J_2 J_3+J_1 J_2+\left(2
   J_1-J_2-J_3\right) \varepsilon +\left(2 J_2 J_3-J_1 \left(J_2+J_3\right)\right)
   \sigma \right)
  \;.
\end{equation}

The explicit form of the coefficients $g_2$ and $g_3$ is  more complicated:
\begin{eqnarray}\nonumber
 g_2&=&\frac{1}{3} \left(J_1^2 \left(J_2^2 (\sigma +1) (\sigma +3)+J_3^2 (\sigma +1) (\sigma
   +3)-J_3 (\sigma +3) \epsilon -J_2 \left(J_3 \sigma  (\sigma +2)+(\sigma +3) \epsilon
   \right)+\epsilon ^2\right)\right.\\
   \nonumber
   &&\left.-J_1 \left(J_2^3 (\sigma +2)+J_2 \left(J_3^2 \sigma
   (\sigma +2)-6 J_3 (\sigma +2) \epsilon +\epsilon ^2\right)\right.\right.\\
   \nonumber
   &&\left.+J_2^2 \left(J_3 \sigma
   (\sigma +2)+(\sigma +3) \epsilon \right)+J_3 \left(J_3+\epsilon \right) \left(J_3
   (\sigma +2)+\epsilon \right)\right)\\
   \nonumber
   &&\left.+J_2^2 \left(J_3^2 (\sigma +1) (\sigma +3)-J_3
   (\sigma +3) \epsilon +\epsilon ^2\right)\right.\\
   \nonumber
   &&\left.-J_1^3 \left(J_2 (\sigma +2)+J_3 (\sigma
   +2)-2 \epsilon \right)\right.\\
   \nonumber
   &&\left.-J_2^3 \left(J_3 (\sigma +2)-2 \epsilon \right)-J_2 J_3
   \left(J_3+\epsilon \right) \left(J_3 (\sigma +2)+\epsilon \right)+J_3^2
   \left(J_3+\epsilon \right){}^2+J_1^4+J_2^4\right),\\
   \label{g2} &&
\end{eqnarray}
and
\begin{equation}\label{g3a}
 g_3=\frac{1}{108}\sum_{i=1}^{8}g_3^{(i)}
 \;,
\end{equation}
where
\begin{eqnarray}
\label{g31}
 g_3^{(1)} &=& 4 J_1^6+ 4 J_2^6 +4 J_3^3 \left(J_3+\epsilon \right){}^3
  -6 J_2^5 \left(J_3 (\sigma +2)-2 \epsilon \right) -6 J_2 J_3^2 \left(J_3+\epsilon \right){}^2 \left(J_3 (\sigma +2)+\epsilon \right),  \\
  \label{g32}
 g_3^{(2)} &=& -6 J_1^5 \left(J_2 (\sigma +2)+J_3 (\sigma +2)-2 \epsilon \right)+
  3 J_2^4 \left(J_3^2 (\sigma  (7 \sigma +10)-1)-2 J_3 (2 \sigma +5) \epsilon +4 \epsilon^2\right),  \\
  \label{g33}
 g_3^{(3)} &=& 3 J_2^2 J_3 \left(J_3^3 (\sigma  (7 \sigma +10)-1)+4 J_3 (2 \sigma +3) \epsilon ^2-2
   J_3^2 \left(\sigma ^2-2\right) \epsilon -2 \epsilon ^3\right),   \\
 \nonumber
 g_3^{(4)} &=&  3 J_1^4 \left(J_2^2 (\sigma  (7 \sigma +10)-1)+J_3^2 (\sigma  (7 \sigma +10)-1)\right.\\
  \label{g34}
 &&\left.-2 J_2
   \left(J_3 \left(5 \sigma ^2-11\right)+(2 \sigma +5) \epsilon \right)-2 J_3 (2 \sigma
   +5) \epsilon +4 \epsilon ^2\right), \\
  \label{g35}
 g_3^{(5)} &=&  2 J_2^3 \left(J_3^3 (\sigma  (\sigma  (2 \sigma -15)-18)+13)-3 J_3 (\sigma +4) \epsilon
   ^2-3 J_3^2 \left(\sigma ^2-2\right) \epsilon +2 \epsilon ^3\right) \\
 \nonumber
 g_3^{(6)} &=&2 J_1^3 \left(J_2^3 (\sigma  (\sigma  (2 \sigma -15)-18)+13)+J_3^3 (\sigma  (\sigma  (2
   \sigma -15)-18)+13)-3 J_3 (\sigma +4) \epsilon ^2\right.\\
   \nonumber
   &&-3 J_2 \left(J_3^2 (\sigma
   ((\sigma -1) \sigma +4)+11)\right.\\
    \label{g36}
  &&\left.\left.
   -4 J_3 (\sigma +2)^2 \epsilon +(\sigma +4) \epsilon
   ^2\right)
      -3 J_3^2 \left(\sigma ^2-2\right) \epsilon -3 J_2^2 \left(J_3 (\sigma
   ((\sigma -1) \sigma +4)+11)+\left(\sigma ^2-2\right) \epsilon \right)+2 \epsilon
   ^3\right),   \\
 \nonumber
 g_3^{(7)} &=& -6 J_1 \left(J_2^5 (\sigma +2)+J_2^3 \left(J_3^2 (\sigma  ((\sigma -1) \sigma +4)+11)-4
   J_3 (\sigma +2)^2 \epsilon +(\sigma +4) \epsilon ^2\right)+J_3 J_2 \left(J_3^3
   \left(5 \sigma ^2-11\right)\right.\right.\\
   \nonumber
   && \left.\left.
   +2 J_3 \sigma  \epsilon ^2-4 J_3^2 (\sigma +2)^2 \epsilon
   -4 \epsilon ^3\right)+J_2^2 \left(J_3^3 (\sigma  ((\sigma -1) \sigma +4)+11)+2 J_3
   \sigma  \epsilon ^2+2 J_3^2 (\sigma  (\sigma +6)+6) \epsilon +\epsilon
   ^3\right)\right.\\
  \label{g37}
   &&\left.+J_2^4 \left(J_3 \left(5 \sigma ^2-11\right)+(2 \sigma +5) \epsilon
   \right)+J_3^2 \left(J_3+\epsilon \right){}^2 \left(J_3 (\sigma +2)+\epsilon
   \right)\right),  \\
    \nonumber
 g_3^{(8)} &=&   3 J_1^2 \left(J_2^4 (\sigma  (7 \sigma +10)-1)+2 J_2^2 \left(J_3^2 (\sigma  (\sigma  (4
   \sigma +3)+18)+33)-2 J_3 (\sigma  (\sigma +6)+6) \epsilon +2 (2 \sigma +3) \epsilon
   ^2\right)\right.\\
   \nonumber
   &&\left.
   +J_3 \left(J_3^3 (\sigma  (7 \sigma +10)-1)+4 J_3 (2 \sigma +3) \epsilon
   ^2-2 J_3^2 \left(\sigma ^2-2\right) \epsilon -2 \epsilon ^3\right)\right.\\
   \nonumber
   &&\left.
   -2 J_2 \left(J_3^3
   (\sigma  ((\sigma -1) \sigma +4)+11)+2 J_3 \sigma  \epsilon ^2+2 J_3^2 (\sigma
   (\sigma +6)+6) \epsilon +\epsilon ^3\right)\right.\\
    \label{g38}
    &&\left.
   -2 J_2^3 \left(J_3 (\sigma  ((\sigma -1)
   \sigma +4)+11)+\left(\sigma ^2-2\right) \epsilon \right)\right)
\;.
\end{eqnarray}

For statistical considerations parts of the phase space with zero measure can be neglected
(but note that the dos may diverge for states with aperiodic motion according to (\ref{Dse})).
Hence we can restrict
ourselves to the ``generic case" where certain exceptions are excluded, see \cite{S21}. In this generic case
the polynomial $\Pi(x)$ will have three real simple roots $x_1<x_2<x_3$  satisfying $x_1+x_2+x_3=0$ and $\Pi(x)>0$
for $x_1<x<x_2$. The explicit form of the roots is known but of overwhelming complexity if expressed in terms
of the physical parameters $\varepsilon,\sigma,J_1,J_2,J_3$.

It follows \cite{S21} that in the generic case (\ref{xd2}) has the solution
\begin{equation}\label{solxt}
 x(t)= \wp \left(t+t_0;g_2,g_3\right)
 \;,
\end{equation}
with the above-mentioned parameters $g_2,g_3$ and the imaginary parameter $t_0$ can be expressed through an
elliptical integral:
\begin{equation}\label{period2}
t_0:={\sf i}\,\int_{x_2}^{x_3}\frac{dx'}{\sqrt{ \left|4 x'^3-g_2 x'-g_3\right|}}
={\sf i}\,\int_{-\infty}^{x_1}\frac{dx'}{\sqrt{\left| 4 x'^3-g_2 x'-g_3\right|}}
=\frac{\sf i}{\sqrt{x_3-x_1}}\,K\left(\frac{x_3-x_2}{x_3-x_1}\right)
\;,
\end{equation}
see  \cite[23.6.34-35]{NIST21} and \cite[17.4.61 ff]{AS72}.
Moreover, this solution will be ${\sf T}$-periodic where
\begin{equation}\label{period1}
\frac{\sf T}{2}=\int_{x_1}^{x_2}\frac{dx'}{\sqrt{ 4 x'^3-g_2 x'-g_3}}=\int_{x_3}^{\infty}\frac{dx'}{\sqrt{ 4 x'^3-g_2 x'-g_3}}
=\frac{1}{\sqrt{x_3-x_1}}\,K\left(\frac{x_2-x_1}{x_3-x_1}\right)
\;.
\end{equation}
Hence, for given coupling constants $J_1,J_2,J_3$, the period ${\sf T}$ can be viewed as a function ${\sf T}(\sigma,\varepsilon)$
although the explicit form of this function is too complicated to be reproduced here.

\section{Distribution of random variables with a saddle point}\label{sec:SP}

\begin{figure}[htb]
\centering
 \includegraphics[width=0.7 \linewidth ]{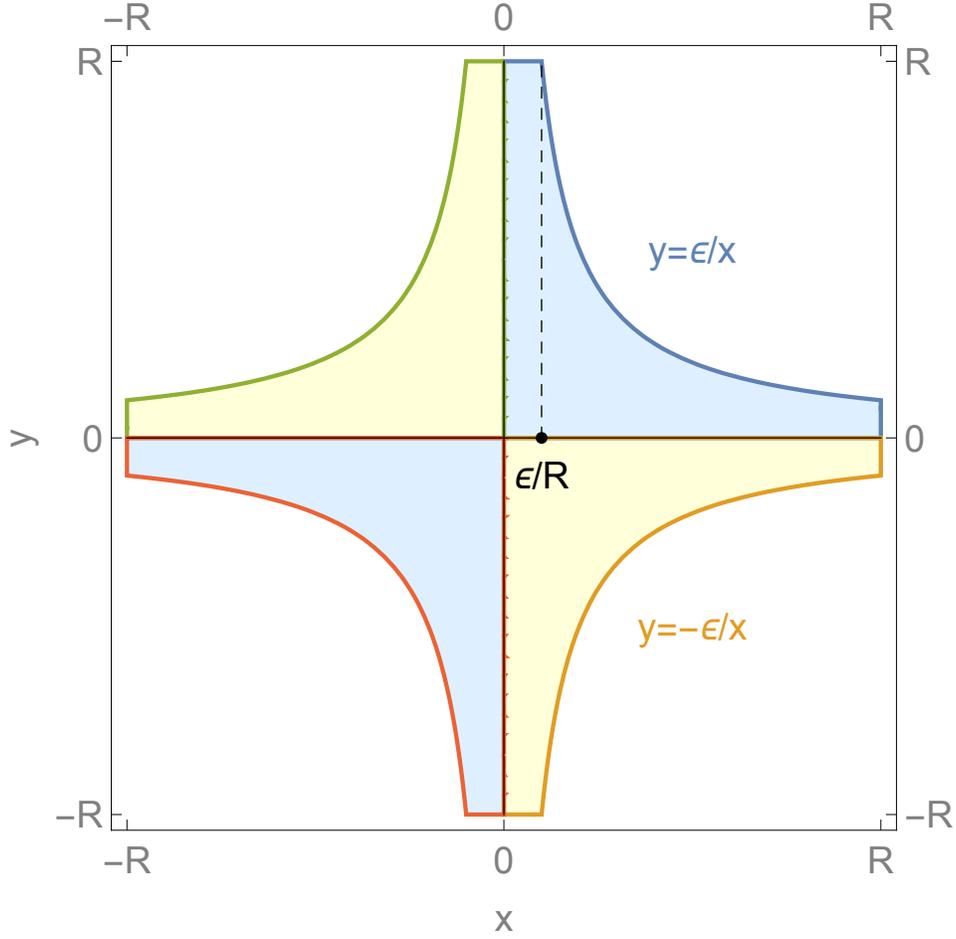}
 \caption{
Plot of the hyperbolic domain $H(\epsilon)$ given by $|x\,y| \le  \epsilon$ and $|x|, \, |y| \le R$.
The area of the intersection of $H(\epsilon)$ with the positive quadrant is given by $\frac{\epsilon}{R}R+\int_{\epsilon/R}^{R}\frac{\epsilon}{x}\,dx $.
}
\label{FIGHEPS}
\end{figure}

We will prove the following Proposition which is tailored to its application in Section \ref{sec:HTP} and not formulated
as general as possible:
\begin{prop}\label{Prop1}
Let ${\mathcal Z}\subset {\mathbbm R}^2$ be an open bounded domain and
$\varphi: {\mathcal Z} \rightarrow {\mathbbm R}_+$ a continuous probability distribution. Further,
let $f: {\mathcal Z}\rightarrow {\mathbbm R}$ be a smooth function (``random variable") with a
saddle point ${\mathbf z}_0\in {\mathcal Z}$ such that $\varphi({\mathbf z}_0)>0$.
Let $\rho^{f}:{\mathbbm R}\rightarrow{\mathbbm R}_+$ be the corresponding probability distribution, i.~e., satisfying
\begin{equation}\label{rhodef}
\int_{u_1}^{u_2}\rho^{f}(u)\,du=\int_{{\mathcal Z}(f;u_1,u_2)}\varphi({\mathbf z})\,d{\mathbf z}
\end{equation}
for all $u_1<u_2\in {\mathbbm R}$ and
\begin{equation}\label{defcalZ}
{\mathcal Z}(f;u_1,u_2):=\{ {\mathbf z}\in {\mathcal Z}\left| \right. u_1\le f({\mathbf z})\le u_2\}
\;.
\end{equation}

Then $\rho^{f}$ has a logarithmic singuarity at $u=u_0:=f\left({\mathbf z}_0\right)$.
\end{prop}

{\bf Proof}: Without loss of generality we may assume $u_0=f\left({\mathbf z}_0\right)=0$.
Then there exist local coordinates $x,y$ in a neighbourhood of ${\mathbf z}_0$ such that
${\mathbf z}_0$ has the coordinates $(0,0)$ and
$f(x,y)=x^2-y^2$ or, after a rotation with $\pi/4$,  $f(x,y)=x\,y$ for, say, $|x|,|y|\le R$ and some $R>0$.
Consider an arbitrary $\epsilon>0$ and choose $u_2=-u_1=\epsilon$ such that
\begin{equation}\label{intHeps}
  p(\epsilon):= \int_{-\epsilon}^{\epsilon}\rho^f(u)\,du = \int_{{\mathcal Z}(f;-\epsilon,\epsilon)}\varphi({\mathbf z})\,d{\mathbf z}
  \ge \int_{H(\epsilon)}\varphi(x,y)\,dx\,dy
  \;,
\end{equation}
where $H(\epsilon)$ is the hyperbolic region
\begin{equation}\label{defHeps}
H(\epsilon):= \{ (x,y)\in{\mathbbm R}^2\left|\right. |x|,|y|\le R,\mbox{ and } \left| x\,y\right| \le \epsilon\}
\;,
\end{equation}
see Figure \ref{FIGHEPS}. By assumption, $\varphi(0,0)>0$ and, since $\varphi$ is continuous, we may choose $R>0$ so small
such that
\begin{equation}\label{boundphi}
 \varphi(x,y)\ge  c >0 \quad \mbox{for all } (x,y)\in H(\epsilon)
 \;.
\end{equation}
This implies
\begin{equation}\label{boundp}
 p(\epsilon)\stackrel{(\ref{intHeps})}{\ge} \int_{H(\epsilon)}\varphi(x,y)\,dx\,dy \ge c\,\left| H(\epsilon)\right|
 \;,
\end{equation}
where $\left|H(\epsilon)\right|$ denotes the area of $H(\epsilon)$ given by
\begin{equation}\label{intarea}
 \left|H(\epsilon)\right| = 4\left(\frac{\epsilon}{R}R+\int_{\epsilon/R}^{R}\frac{\epsilon}{x}\,dx \right)
 =
 4\left( \epsilon- \epsilon \log\epsilon+ 2\epsilon \log R\right)
 \;,
\end{equation}
see Figure \ref{FIGHEPS}. If $u\mapsto\rho^f(u)$ would be continuous in a neighbourhood of $u=0$ then it would follow that
\begin{equation}\label{limrho0}
\rho^f(0) \stackrel{(\ref{intHeps})}{=} \lim_{\epsilon\to 0}\frac{p(\epsilon)}{2\epsilon}
\stackrel{(\ref{boundp},\ref{intarea})}{\ge}  \lim_{\epsilon\to 0} 2c \left( 1-\log\epsilon+2\log R\right)
\;,
\end{equation}
which is a contradiction due to the divergence of $-\log \epsilon$. Hence $\rho^f(0)$ is divergent
and the singularity is, at least, of logarithmic order.

The singularity is exactly of logarithmic order since the contribution to $\rho^f(0)$ from other possible zeroes $f\left({\mathbf z}_\nu\right)=0$
except the considered saddle point at ${\mathbf z}_0$ would be of order $O(1)$ for regular zeroes, or of order $O(\epsilon)$ for
local maxima or minima, or of order $O(\log\epsilon)$ for other saddle points.           \hfill$\Box$\\

\section{ Asymptotic expansions of Fourier integrals involving logarithmic singularities}\label{sec:ALS}

We consider the case of a distribution function $\rho(\omega)$ with a logarithmic singularity at $\omega=\omega_0$
and will investigate the decay of the corresponding Fourier transform $\widehat{\rho}(t)$ for $t\to\infty$. More specifically, we assume that
$\rho(\omega)$  is of the form
\begin{equation}\label{formofrho}
 \rho(\omega)=\phi(\omega)\,\log\left| \omega-\omega_0\right|
 \;,
\end{equation}
where $\phi(\omega)$ is $N$ times continuously differentiable for $\gamma_1< \omega < \gamma_2$ and $\gamma_1< \omega_0 < \gamma_2$.
For our purposes we may assume that $\phi(\omega)$ is a real function.
Then we consider the Fourier integral
\begin{equation}\label{Fourieromega}
 \widehat{\rho}(t) = \int_{\gamma_1}^{\gamma_2}\phi(\omega)\,\log\left| \omega-\omega_0\right|\,\exp({\sf i}\,\omega\,t)\,d\omega
 \;,
\end{equation}
and the asymptotic expansion of $\widehat{\rho}(t)$ for $t\to \infty$. This problem has been solved in \cite[Th.~4]{E56}
for the \underline{one-sided} Fourier integral
\begin{equation}\label{Erd1}
 A= \int_{\omega_0}^{\gamma_2}\phi(\omega)\,\log\left| \omega-\omega_0\right|\,\exp({\sf i}\,\omega\,t)\,d\omega
 \;.
\end{equation}
We will utilize this solution to obtain the asymptotic expansion for the \underline{two-sided} Fourier integral (\ref{Fourieromega}).
For this purpose, we will quote the corresponding theorem $4$ of \cite{E56} in full detail, with slight modifications according to our notation.
\begin{prop}\label{PropErdelyi}(Erd\'{e}lyi)
 Under the preceding assumptions on $\phi(\omega)$ we have
 \begin{equation}\label{asymptoticErdelyi}
  A=\sum_{n=0}^{N-1} {\sf i}^{n+1}\,\phi^{(n)}(\omega_0)\left[\psi(n+1)-\log t+{\sf i}\frac{\pi}{2} \right]\,t^{-n-1}\,\exp\left({\sf i}\,\omega_0 t \right)
  +o\left(t^{-N}\right)
  \;,
 \end{equation}
 for $t\to +\infty$, where $\psi(z)$ denotes the logarithmic derivative of $\Gamma(z)$.
\end{prop}
Let us denote the complementary integral of (\ref{Erd1}) by
\begin{equation}\label{Erd2}
 \widetilde{A}=  \int_{\gamma_1}^{\omega_0}\phi(\omega)\,\log\left| \omega-\omega_0\right|\,\exp({\sf i}\,\omega\,t)\,d\omega
 \;,
\end{equation}
such that $\widehat{\rho}(t) =A+\widetilde{A}$ and denote complex conjugation by an overline. Then
\begin{eqnarray}
\label{Erd3a}
  \overline{\widetilde{A}} &=& \overline{ \int_{\gamma_1}^{\omega_0}\phi(\omega)\,\log\left| \omega-\omega_0\right|\,\exp({\sf i}\,\omega\,t)\,d\omega }\\
  \label{Erd3b}
   &=&  \overline{ -\int_{-\gamma_1}^{-\omega_0}\phi(-\omega)\,\log\left| \omega-\omega_0\right|\,\exp(-{\sf i}\,\omega\,t)\,d\omega} \\
   \label{Erd3c}
   &=& \int_{-\omega_0}^{-\gamma_1}\phi(-\omega)\,\log\left| \omega-\omega_0\right|\,\exp({\sf i}\,\omega\,t)\,d\omega
   \;,
\end{eqnarray}
where we have used that $\phi(\omega)$ is real. This form of $ \overline{\widetilde{A}}$ is suited for the application of Proposition \ref{PropErdelyi}.
Using the abbreviation $\phi(-\omega)=\widetilde{\phi}(\omega)$ which yields
\begin{equation}\label{phin}
 \widetilde{\phi}^{(n)}\left(-\omega_0\right)=\left(-1\right)^n \,\phi^{(n)}\left( \omega_0\right)\,d\omega
\end{equation}
we thus obtain from (\ref{asymptoticErdelyi}) and the replacement $\omega_0\mapsto -\omega_0$
\begin{equation}\label{Erd4}
  \overline{\widetilde{A}}=\sum_{n=0}^{N-1} {\sf i}^{n+1}\,\widetilde{\phi}^{(n)}(-\omega_0)\left[\psi(n+1)-\log t+{\sf i}\frac{\pi}{2} \right]\,t^{-n-1}\,\exp\left(-{\sf i}\,\omega_0 \,t \right)
  +o\left(t^{-N}\right)
  \;.
\end{equation}
This entails
\begin{eqnarray}
\label{Erd5a}
 \widetilde{A} &\stackrel{(\ref{phin})}{=}&\sum_{n=0}^{N-1}\left( -{\sf i}\right)^{n+1}\,(-1)^n\,{\phi}^{(n)}(\omega_0)
 \left[\psi(n+1)-\log t-{\sf i}\frac{\pi}{2} \right]\,t^{-n-1}\,\exp\left({\sf i}\,\omega_0\, t \right)
  +o\left(t^{-N}\right) \\
  \label{Erd5b}
   &=& -\sum_{n=0}^{N-1} {\sf i}^{n+1}\,{\phi}^{(n)}(\omega_0)
 \left[\psi(n+1)-\log t-{\sf i}\frac{\pi}{2} \right]\,t^{-n-1}\,\exp\left({\sf i}\,\omega_0\, t \right)
  +o\left(t^{-N}\right)
  \;,
\end{eqnarray}
and, finally,
\begin{equation}\label{Erd6}
 \widehat{\rho}(t)= A+\widetilde{A}=-\pi\sum_{n=0}^{N-1}{\sf i}^n \,\phi^{(n)}\left(\omega_0\right)\,t^{-n-1}\,\exp\left({\sf i}\,\omega_0\, t \right)
  +o\left(t^{-N}\right)
  \;.
\end{equation}
We note that the terms containing $\log t$ cancel
and the leading term corresponding to $n=0$ in the asymptotic expansion (\ref{Erd6}) is proportional to $1/t$.


\end{document}